\documentclass[12pt]{article}
\usepackage{amsmath}
\usepackage{graphicx}
\usepackage{natbib}
\usepackage{url} 

\newcommand{\blind}{0}

\usepackage{xr}
\RequirePackage{amsmath,amsfonts,amssymb}
\RequirePackage{graphicx}
\usepackage{slashbox}
\usepackage{subfigure,epsfig,psfrag}
\usepackage{verbatim,float,url,enumerate}
\usepackage{bm,dsfont,color,appendix}
\usepackage{adjustbox}
\usepackage{mathtools}
\usepackage{bbm}
 \usepackage{booktabs}
\usepackage{latexsym,graphicx}
\usepackage{tikz}
\usetikzlibrary{arrows,positioning}
\usepackage{rotating}

\usepackage[linesnumbered,ruled]{algorithm2e}

\newtheorem{theorem}{Theorem}[section]
\newtheorem{lemma}[theorem]{Lemma}
\newtheorem{proposition}[theorem]{Proposition}
\newtheorem{remark}{Remark}[theorem]

\usepackage{titlesec}
\titlespacing*{\section}{0pt}{0\baselineskip}{0\baselineskip}
\titlespacing*{\subsection}{0pt}{0\baselineskip}{0\baselineskip}
\titlespacing*{\subsubsection}{0pt}{0\baselineskip}{0\baselineskip}

\def\deta{{\boldsymbol \eta}}
\def\bz{{\bf z}}
\def\by{{\bf y}}
\def\bY{{\bf Y}}

\def\bW{{\bf W}}

\def\real{{\mathbb R}}
\def\boldf{{\bf f}}

\def\zero{{\boldsymbol 0}}
\def\b1{{\boldsymbol 1}}

\def\diag{\mathrm{diag}}

\def\bP{\bold P}
\def\bF{\bold F}
\def\bH{\bold H}
\def\bSigma{\bold \Sigma}

\def\bbeta{\boldsymbol{\beta}}
\def\bmu{\boldsymbol{\mu}}
\def\bB{\boldsymbol{B}}
\def\bA{\boldsymbol{A}}
\def\bC{\boldsymbol{C}}
\def\bX{\boldsymbol{X}}

\def\bv{\boldsymbol{v}}
\def\bu{\boldsymbol{u}}
\def\bmu{\boldsymbol{\mu}}
\def\bU{\boldsymbol{U}}
\def\bM{\boldsymbol{M}}
\def\bG{\boldsymbol{G}}
\def\bV{\boldsymbol{V}}
\def\bE{\mathbb{E}}
\def\bZ{\boldsymbol{Z}}
\def\Id{{\rm {\mathbf Id}}}
\def\bPhi{\boldsymbol{\Phi}}
\def\bphi{\boldsymbol{\phi}}
\usepackage{xr-hyper} 
\begin{document}

\def\spacingset#1{\renewcommand{\baselinestretch}%
{#1}\small\normalsize} \spacingset{1}


\if0\blind
{
  \title{\bf Smooth and probabilistic PARAFAC model with auxiliary covariates}
  \author{Leying Guan\hspace{.2cm}\\
    Department of Biostatistics, Yale University}
  \maketitle
} \fi

\if1\blind
{
  \bigskip
  \bigskip
  \bigskip
  \begin{center}
    {\LARGE\bf Smooth and probabilistic PARAFAC model with auxiliary covariates}
\end{center}
  \medskip
} \fi

\bigskip

\begin{abstract}
In immunological and clinical studies, matrix-valued time-series data clustering is increasingly popular. Researchers are interested in finding low-dimensional embedding of subjects based on potentially high-dimensional longitudinal features and investigating relationships between static clinical covariates and the embedding. These studies are often challenging due to high dimensionality, as well as the sparse and irregular nature of sample collection along the time dimension. We propose a smoothed probabilistic PARAFAC model with covariates (SPACO)  to tackle these two problems while utilizing auxiliary covariates of interest.   We provide intensive simulations to test different aspects of SPACO and demonstrate its use on an immunological data set from patients with SARs-CoV-2 infection.
\end{abstract}

\noindent%
\vfill

\newpage
\spacingset{1.5} 
\section{Introduction}
\label{sec:intro}
Sparsely observed multivariate times serious data are now common in immunological studies. For each subject or participant i $(i=1,\ldots, I)$, we can collect multiple measurements on $J$ features over time, but often at $n_i$ different time stamps $\{t_{i,1},\ldots, t_{i,n_i}\}$. For example, for each subject, immune profiles are measured for hundreds of markers at irregular sampling times in \cite{lucas2020longitudinal} and \cite{rendeiro2020longitudinal}. Let $\bX_i\in \mathbb{R}^{n_i\times J}$ be the longitudinal measurements for subject $i$, we can collect $\bX_i$ for all $I$ subjects into a sparse three-way tensor $\bX\in \real^{I\times T\times J}$, where $T=|\cup_{i}\{t_{i, 1}, \ldots, t_{i,n_i}\}|$ is the number of distinct time stamps across all subjects.   Since  \{$t_{i, 1}, \ldots, t_{i,
n_i}$\} tend to be small in size and have low overlaps for different subject $i$, $\bX$ may have a high missing rate along the time dimension.

In addition to $\bX_i$, researchers often have a set of nontemporal covariates $\bz_i\in \mathbb{R}^{q}$ for subject $i$ such as medical conditions and demographics, which may account partially for the variation in the temporal measurements $\bX$ across subjects. Modeling such auxiliary covariates $\bZ\coloneqq (\bz_1, \ldots, \bz_I)^\top$ together with $\bX$ might help with the estimation quality and understanding for the cross-subject heterogeneity. 

In this paper, we propose SPACO (\underline{s}mooth and probabilistic  \underline{PA}RAFAC model with auxiliary \underline{co}variates) to adapt to the sparsity long the time dimension in $\bX$ and  utilize the auxiliary
variables $\bZ$. SPACO assumes that  $\bX$ is a noisy realization of some low-rank signal tensor whose time components are smooth and  subject scores have a potential dependence on the auxiliary covariates $\bZ$:
\begin{align}
\label{eq:eq_obs}
&x_{itj}=\sum_{k=1}^K u_{ik}\phi_{tk}v_{jk}+\epsilon_{itj},\quad 
\epsilon_{itj}\sim \mathcal{N}(0, \sigma^2_j)\nonumber\\
&\bu_{i}=(u_{ik})_{k=1}^K\sim \mathcal{N}(\deta_i, \Lambda_f),
\quad \deta_i = \bbeta^\top\bz_i.\nonumber
\end{align}
Here, (1) $\bbeta\in \mathbb{R}^{q\times K}$ describes the
dependence of the expected subject score $\beta_i$ for subject $i$ on $\bz_i$, and (2) $u_{ik}$, $\phi_{tk}$, $v_{jk}$ are the subject score, trajectory
value and feature loading for factor $k$ in the PARAFAC model  and the observation indexed by $(i,t,j)$ where $\bu_i$ has a normal prior $\mathcal{N}(\deta_i, \Lambda_f)$.  We impose smoothness
on time trajectories $(\phi_{tk})_{t=1}^\top$ and sparsity on $\bbeta$ to deal with the irregular sampling along the time dimension and potentially high dimensionality in $\bZ$. 

Alongside the model proposal, we will also discuss several issues related to SPACO, including model initialization,  auto-tuning of smoothness and sparsity in $\beta$ and hypothesis testing on $\beta$ through cross-fit. Successfully addressing these issues is crucial to practitioners interested in applying SPACO to their analysis.

In the remaining of the article, we summarize some closely related work in section \ref{sec:related} and describe the SPACO model  in Section \ref{sec:model} and model parameter estimation with fixed tuning parameters in Section \ref{sec:estimation}. In Section \ref{sec:extensions}, we discuss the aforementioned related issues that could be important in practice. We compare SPACO to several existing methods in Section \ref{sec:sims}. Finally, in Section \ref{sec:case_study}, we apply SPACO to a highly sparse tensor data set on immunological measurements for SARS-COV-2 infected patients. We provide a python package {\it SPACO} for researchers interested in applying the proposed method.

\subsection{Related work}
\label{sec:related}
In the study of multivariate longitudinal data in economics, researchers have combined tensor decomposition with auto-cross-covariance estimation and autoregressive models \citep{fan2008high,lam2011estimation,fan2011high, bai2016econometric,  wang2019factor,wang2021high}. These approaches do not
work well with highly sparse data or do not scale well with the feature dimensions, which are important for working with medical data.   Functional PCA \citep{besse1986principal,
yao2005functional} is often used  for modeling sparse longitudinal data in the matrix-form. It utilizes the smoothness of time trajectories to handle sparsity in the longitudinal observations, and estimates the eigenvectors and factor scores under this smoothness assumption. \cite{yokota2016smooth} and \cite{imaizumi2017tensor} introduced smoothness to tensor decomposition models, and estimated the model parameters by iteratively solving penalized regression problems. The methods above don't consider the auxiliary covariates $\bZ$.

It has been previously discussed that including $\bZ$ could potentially improve our estimation.  \cite{li2016supervised} proposed SupSFPC (supervised sparse and functional principal component) and observed that the auxiliary covariates improve the signal estimation quality in the matrix setting for modeling multivariate longitudinal observations.  \cite{lock2018supervised} proposed SupCP which performs supervised multiway factorization model with complete observation and employs a probabilistic tensor model \citep{tipping1999probabilistic,mnih2007probabilistic,hinrich2019probabilistic}. Although an extension to sparse tensor is straightforward, SupCP does not model the smoothness and can be much more affected by severe missingness along the time dimension.

SPACO can be viewed as an extension of functional PCA and SupSFPC  to the setting of three-way tensor decomposition \citep{acar2008unsupervised,sidiropoulos2017tensor} using the parallel factor (PARAFAC) model \citep{harshman1994parafac,carroll1980candelinc}. It uses a probabilistic model and jointly models the smooth longitudinal data
with potentially high-dimensional non-temporal covariates $\bZ$. We refer to the SPACO model as SPACO- when no auxiliary covariate $\bZ$  is available. SPACO- itself is an attractive alternative to existing tensor decomposition implementations with probabilistic modeling, smoothness regularization, and automatic parameter tuning. In our simulations, we compare SPACO with SPACO to demonstrate the gain from utilizing  $\bZ$.

\section{SPACO Model}
\label{sec:model}
\subsection{Notations}
Let $\bX\in \mathbb{R}^{I\times T\times J}$ be a tensor for some sparse
multivariate longitudinal observations, where $I$ is the number of subjects, $J$ is the number of features, and $T$ is the number of total unique time points.
For any matrix $A$, we let $A_{i:}\slash A_{:i}$ denote its $i^{th}$ row\slash
column, and often write $A_{:i}$ as  $A_{i}$ for the $i^{th}$ column for convenience. Let
$\bX_{I}\coloneqq \begin{pmatrix} \bX_{:,:,1} & \cdots & \bX_{:,:, J}
\end{pmatrix}\in \mathbb{R}^{I\times (TJ)}$, $\bX_T\coloneqq \begin{pmatrix}
\bX_{:,:,1}^\top & \cdots &  \bX_{:,:, J}^\top\end{pmatrix} \in
\mathbb{R}^{T\times (IJ)}$,  $\bX_J\coloneqq \begin{pmatrix}
\bX^\top_{:,1,:} &\cdots & \bX_{:,T, :}^\top \end{pmatrix}\in
\mathbb{R}^{J\times (IT)}$  be the matrix unfolding of $\bX$ in the subject\slash feature\slash time dimension respectively. We also define:

\noindent {\textbf{Tensor product $\circledcirc$}}: $a\in \mathbb{R}^{I}$, $b\in \mathbb{R}^{T}$, $c\in \mathbb{R}^{J}$, then, $A = a\circledcirc b\circledcirc c \in \mathbb{R}^{I\times T\times J}$ with $A_{itj}=a_i b_t c_j$.

\noindent{\textbf{Kronecker product $\otimes$}}:  $A \in \mathbb{R}^{I_1\times K_1}$, $B \in \mathbb{R}^{I_2\times K_2}$, then 
\[
C = A\otimes B = \left(\begin{array}{lll}A_{11}B&\ldots&A_{1K_1}B\\\vdots&\ddots&\vdots\\A_{I_1 1}B& \ldots& A_{I_1 K_1}B\end{array} \right) \in \mathbb{R}^{(I_1 I_2)\times (K_1K_2)}.
\]
{\textbf{Column-wise Khatri-Rao product $\odot$}}: $A \in
\mathbb{R}^{I_1\times K}$, $B\in \mathbb{R}^{I_2\times K}$, then $C=A\odot B \in
\mathbb{R}^{(I_1 I_2)\times K}$ with $C_{:,k} = (A_{:,k}\otimes B_{:,k})$ for $k=1,\ldots, K$.

\noindent{\textbf{Element-wise multiplication $\cdot$}}:   $A, B \in \mathbb{R}^{I\times K}$,  then
$C = A\cdot B \in\mathbb{R}^{I\times K}$ with $C_{ik}= (A_{ik}B_{ik})$; for $b \in \mathbb{R}^{K}$, $C = A\cdot b =A\diag\{b_1,\ldots, b_K\}$; for $b \in \mathbb{R}^{I}$, $C = b\cdot A = \diag\{b_1,\ldots, b_I\}A$.
\subsection{smooth and probabilistic  PARAFAC model with  covariates}
We assume $\bX$ to be a noisy realization of an underlying signal array $\bF
=\sum_{k=1}^K \bU_{k}\circledcirc \bPhi_k \circledcirc  \bV_{k}$.
We stack $\bU_k/\bPhi_k/\bV_k$ as the columns of $\bU/\bPhi/\bV$, denote the
rows of $\bU/\bPhi/\bV$ by
$\bu_i\slash \bphi_t\slash\bv_j$, and their entries by
$u_{ik}\slash \phi_{tk}\slash v_{jk}$. We let $x_{itj}$ denote the $(i,t,j)$-entry of $\bX$. Then,
\begin{equation}
\label{eq:model}
x_{itj}=\sum_{k=1}^K u_{ik}\phi_{tk}v_{jk}+\epsilon_{itj},\;\bu_{i}\sim \mathcal{N}(\deta_i, \Lambda_f),\;\epsilon_{ijt}\sim \mathcal{N}(0, \sigma^2_j),
\end{equation}
where $\Lambda_f = \diag\{s_1^2,\ldots,s_K^2\}$ is a
$K\times K$ diagonal  covariance matrix.  Even though $\bX$ is often of high rank, we consider the scenario where the rank $K$ of $\bF$ is small.

Without covariates, we set the prior mean
parameter $\deta_i =\zero$.  If we are interested in explaining the heterogeneity in $\deta_i$ across subjects with auxiliary covariates $\bZ\in\mathbb{R}^{I\times q}$, then we may model $\deta_i$ as a function of $\bz_i\coloneqq \bZ_{i,:}$. Here, we consider a linear model $\eta_{ik} = \bz_i^\top\bbeta_k, \quad \forall k = 1,\ldots, K$. To avoid confusion, we will always call $\bX$ the ``features'', and $\bZ$ the
``covariates'' or ``variables''.

We refer to $\bU$ as the subject scores, which characterize differences across subjects and are latent variables. We refer to $\bV$ as the feature loadings, which reveal the composition of the factors using the original features and could assist the downstream interpretation. Finally, $\bPhi$ is referred to as the time trajectories, which can be interpreted as function values sampled from some underlying smooth functions $\phi_k(t)$ at a set of discrete-time points, e.g., $\bPhi_k = (\phi_k(t_1),\ldots, \phi_k(t_T))$.

Recalling that $\bX_{I} \in \mathbb{R}^{I\times (TJ)}$ is the unfolding of $\bX$
in the subject direction, we write $\vec i$ for the indices of observed values
in the $i^\text{th}$ row of $\bX_I$, and $\bX_{I,\vec i}$ for the vector of
these observed values. Each such observed value $x_{itj}$ has noise variance
$\sigma_j^2$, and we write $\Lambda_{\vec i}$ to represent the diagonal covariance matrix with diagonal values $\sigma_j^2$ being the corresponding variances for $\varepsilon_{ijt}$ at indices in $\vec
i$. Similarly, we define $\{\vec t,\; \bX_{T,\vec t},\;\Lambda_{\vec t}\}$ for
the unfolding $\bX_{T} \in \mathbb{R}^{T\times (IJ)}$, and $\{\vec j,
\;\bX_{J,\vec j},\;\Lambda_{\vec j}\}$ for the observed indices, the associated observed vector and diagonal covariance matrix for the $j^{th}$ row in  $\bX_{J} \in \mathbb{R}^{J\times
(IT)}$. We set $\Theta = \{\boldsymbol{V}, \boldsymbol{\Phi}, {\boldsymbol \beta}, \left(\sigma^2_j, j= 1,\ldots, J\right),
\left(s^2_{k}, k = 1,\ldots, K\right)\}$ to denote all model parameters. Set $\bH = (\bV\odot\bPhi)$ and $\boldf_i= \bX_{I,\vec i}-\bH_{\vec
i}\deta_i$.  If $\bU$ is observed, the complete data log-likelihood is 
\begin{align}
\label{eq:loglik_complete}
L(\bX,\bU|\Theta)=&-\frac{1}{2}\sum_i \left(\boldf_i^\top\Lambda^{-1}_{\vec i} \boldf_i+(\bu_i - \deta_i)^\top\Lambda_f^{-1} (\bu_i - \deta_i)+\log |\Lambda_{\vec i}|+I\log |\Lambda_f|\right).
\end{align}
Set $\tilde\Sigma_{i} =  \Lambda_{\vec i}+\bH_{\vec i}\Lambda_{f} \bH_{\vec i}^{\top}$. The marginalized log likelihood integrating out the randomness in $\bU$ enjoys a closed form \citep{lock2018supervised}:
\begin{align}
\label{eq:loglik_marginal1}
L(\bX|\Theta)&\propto  -\frac{1}{2}\left(\sum_{i}  \boldf_i^\top\tilde\Sigma_{i}^{-1} \boldf_i+\log |\tilde\Sigma_{i}|\right).
\end{align}
Set $\Sigma_i =  (\bH_{\vec i}^\top \Lambda_{\vec i}^{-1} \bH_{\vec i}+\Lambda_{f}^{-1})^{-1}$. We can also equivalent express the marginal likelihood as below.
\begin{align}
\label{eq:loglik_marginal2}
L(\bX|\Theta)&\propto   -\frac{1}{2}\boldf_i^\top\left( \Lambda_{\vec i}^{-1}- \Lambda_{\vec i}^{-1}\bH_{\vec i}\Sigma_i \bH_{\vec i}^\top \Lambda_{\vec i}^{-1}\right)\boldf_i-\frac{1}{2}\left(\log |\Lambda_{\vec i}|+\log |\Lambda_f| - \log |\Sigma_i| \right),
\end{align}
We use the form in eq.\ (\ref{eq:loglik_marginal2}) to derive the updating formulas and criteria for rank selection since it does not involve the inverse of a large non-diagonal matrix.

Model parameters in eq.\ (\ref{eq:loglik_marginal1}) or  eq.\ (\ref{eq:loglik_marginal2}) are not identifiable due to (1) parameters rescaling  from $(\bPhi_k, \bV_k, \bbeta_k, s_k^2)$ to  $(c_1\bPhi_k, c_2\bV_k, c_3\bbeta_k, c_3^2s_k^2)$ for any $c_1 c_2 c_3 = 1$, and (2) reordering of different component $k$ for $k=1,\ldots, K$. More discussions of the model identifiability can be found in \cite{lock2018supervised}. Hence, adopting similar rules from \cite{lock2018supervised}, we require
\begin{itemize}
\item [(C.1)] $\|\bV_k\|_2^2 = 1$, $\|\bPhi_k\|_2^2 = T$.
\item [(C.2)] The latent components are in decreasing order based on their overall variances  $\Lambda_{f,kk}+\|\bZ\bbeta_{k}\|_2^2\slash I$, and the first non-zero entries in $\bV_k$ and $\bPhi_k$ to be positive, e.g., $v_{k1} > 0$ and $\phi_{k1} > 0$ if they are non-zero.
\end{itemize}
To deal with the sparse sampling along the time dimension and take into consideration that features are often smooth over time in practice, we  assume that the time component $\bPhi_k$ is sampled from a slowly varying trajectory function $\phi_k(t)$, and  encourage smoothness of $\phi_k(t)$ by
directly penalizing the function values via a penalty term
$\sum_{k}\lambda_{1k}\bPhi_k^\top\Omega\bPhi_k$.   This paper considers a Laplacian smoothing
\citep{sorkine2004laplacian} with a weighted adjacency matrix $\Gamma$.  Let $\mathcal{T}(t)$ represent the associated time for $\bX_{.,t.,}$. We define $\Omega$ and $\Gamma$ as
\[
\Omega = \Gamma^\top\Gamma,\quad \Gamma = \left(\begin{array}{lllll}\frac{1}{\mathcal{T}[2]-\mathcal{T}[1]}&-\frac{1}{\mathcal{T}[2]-\mathcal{T}[1]}&\ldots&0&0\\
0&\frac{1}{\mathcal{T}[3]-\mathcal{T}[2]}&\ddots&0&\vdots\\
\vdots&\vdots&\ddots&\vdots&\vdots\\
0&0&\ldots&\frac{1}{\mathcal{T}[T]-\mathcal{T}[T-1]}&-\frac{1}{\mathcal{T}[T]-\mathcal{T}[T-1]}
\end{array}\right)\in \mathbb{R}^{T\times(T-1)}
\]
Practitioners may choose other forms for $\Omega$. If practitioners want $\phi_k(t)$ to have slowly varying derivatives, they can also use a penalty matrix that penalizes changes in gradients over time.

Further, when the number of covariates $q$ in $\bZ$ is moderately large, we may wish to impose sparsity in the $\bbeta$ parameter. We encourage such sparsity by including a lasso penalty \citep{tibshirani2011regression} in the model. In summary, our goal is then to find parameters maximizing the expected penalized log-likelihood, or minimizing the penalized expected deviance loss, under norm constraints:
\begin{align}
\label{eq:objective}
\min&\; J(\Theta)\coloneqq-\frac{1}{2} L(\bX|\Theta)+ \sum_{k=1}^K
\lambda_{1k}\bPhi_k^\top\Omega \bPhi_k+\sum_k\lambda_{2k} |\bbeta_k|\nonumber\\
&\quad  \text{s.t} \quad \|\bV_k\|_2^2 = 1,\;\|\bPhi_k\|_2^2=T, \text{ for all } k = 1,\ldots, K.
\end{align}
Only the identifiability constraint (C.1) has entered the objective. We can always guarantee (C.2) by changing the signs in $\bV$, $\bPhi$, $\bbeta$ and reordering the components afterward without changing the achieved objective value.

Eq.\ (\ref{eq:objective}) describes a non-convex problem.  We will find locally optimal solutions via an alternating update procedure: (1) fixing other parameters and updating $\bbeta$ via lasso regressions; (2) fixing $\bbeta$ and updating other model parameters using the EM algorithm. We give details of our iterative estimation procedure in Section \ref{sec:estimation}.

\section{Model parameter estimation}
\label{sec:estimation}
Given the model rank $K$ and penalty terms $\lambda_{1k}$, $\lambda_{2k}$, we alternately update
parameters $\bbeta$, $\bV$, $\bU$, $s^2$ and $\sigma^2$ with a mixed EM
procedure described in Algorithm \ref{alg:basicAlg}. We briefly explain the
updating steps here:

\noindent\textbf{(1):} Given other parameters, we find $\bbeta$ to to directly minimize the objective  by solving a
least-squares regression problem with lasso penalty.

\noindent\textbf{(2):}  Fixing $\bbeta$, we update the other parameters using an EM procedure.
Denote the current parameters as $\Theta_0$.  Our goal is to minimize the
penalized expected negative log-likelihood
\begin{equation}
\label{eq:EMobj}
J(\Theta; \Theta_0)\coloneqq \bE_{\bU|\Theta_0}\left\{-L(\bX,\bU|\Theta)+\sum_k \lambda_{1k}\Phi_k^\top\Omega\Phi_k+\sum_{k}\lambda_{2k}|\bbeta_k|\right\},
\end{equation}
under the current posterior distribution $\bU|\Theta_0$. We adopt a block-wise parameter updating scheme where we update $\bV_k$, $\bPhi_k$, $\Lambda_f$ and $\sigma_j^2$ sequentially. 

\begin{algorithm}[H]
\caption{SPACO with fixed penalties}
\label{alg:basicAlg}
\KwData{$\bX$, $\Omega$, $\lambda_1$, $\lambda_2$, $K$}
\KwResult{Estimated $\bV$, $\bPhi$, $\bbeta$, $s^2$, $\sigma^2$ and  posterior $\bP(\bU|\Theta, \bX)$ and the marginalized density $\bP(\bX|\Theta)$ .}

Initialization of  $\bV$, $\bPhi$, $\bbeta$, $s^2$, $\sigma^2$ and the posterior distribution of $\bU$\;

\While{Not converged}{
\For{$k=1,\ldots, K$}{
  $	\bbeta_{:,k}\leftarrow \arg\min_{\beta_{:,k}} \{-L(\bX|\Theta)+\lambda_{2k}|\bbeta_{:,k}|\}$

  $\bV_k \leftarrow \arg\min_{\bV_k}\left[J(\Theta; \Theta_0)+\nu \|\bV_k\|_2^2\right]$, $\nu$ is the largest value leading to the minimizer having $\|\bV_k\|_2^2=1$
  
  $\bPhi_k \leftarrow \arg\min_{\bV_k}\left[J(\Theta; \Theta_0)+\nu \|\bU_k\|_2^2\right]$, $\nu$ is the largest value leading to the minimizer having $\|\bPhi_k\|_2^2=T$

   $ s^2_k\leftarrow \arg\min_{s^2_k} \bE_{\bU|\Theta_0} L(\bX, \bU|\Theta)$.

}

For $j = 1,\ldots, J$: $\sigma^2_j\leftarrow \arg\min_{\sigma^2_j}\bE_{\bU|\Theta_0} L(\bX, \bU|\Theta)$.

Update the posterior distribution of $\bU$.
}
\vspace{-.5em}
\end{algorithm}
Algorithm \ref{alg:basicAlg} describes the high level ideas of our updating schemes. In line 5 and 6, we guarantee the norm constraints on $\bV_k$ and $\bPhi_k$ by adding an additional quadratic term and set the coefficient $\nu$ to guarantee the norm requirements. Even though this is not a convex problem, the proposed approaches provide optimal solutions for sub-routines updating different parameter blocks, and the penalized (marginalized) deviance loss is non-increasing over the iterations.


\begin{theorem}
\label{thm:monotone}
In Algorithm \ref{alg:basicAlg}, let $\Theta_{\ell}$ and $\Theta_{\ell+1}$ are the estimated parameters at the beginning and end of the $\ell^{th}$ iteration of the outer \textbf{while} loop. We have $J(\Theta^{\ell})\geq J(\Theta^{\ell+1})$.
\end{theorem}
Proof of Theorem \ref{thm:monotone} is given in Appendix \ref{app:proof_monotone}. In Algorithm \ref{alg:basicAlg}, the posterior distribution of $\bu_i$ for each row in $\bU$ is Gaussian, with posterior
covariance the same as $\Sigma_i$ defined earlier, and posterior  mean  given below.
\begin{align}
\label{eq:Uposterior}
 \bmu_i =\Sigma_i\left(\Lambda_f^{-1}\bbeta^\top \bz_i+(\bH)_{\vec
i}^\top\Lambda^{-1}_{\vec i} \bX_{I,\vec i}\right).
\end{align}
Explicit formulas and steps for carrying out the subroutines at lines 4-7 and line 9 are deferred to Appendix \ref{app:alg_basic}
\section{Initialization,  tuning and testing}
\label{sec:extensions}
\subsection{Initialization}
One initialization approach is to form a Tucker decomposition $[\bU_{\perp}, \bPhi_{\perp}, \bV_{\perp}; \bG]$ of $\bX$ using HOSVD\slash MLSVD \citep{de2000multilinear} where $\bG\in \real^{K_1\times K_2\times K_3}$ is the core tensor and $\bU_{\perp}\in \real^{I\times K_1}$, $\bPhi_{\perp}\in\real^{T\times K_2}$, $\bV_{\perp}\in \real^{J\times K_3}$ are unitary matrices multiplied with the core tensors along the subject,  time and feature directions  respectively ($K_1\slash K_2\slash K_3$ is the smallest between $K$ and $I\slash T\slash J$), and then perform PARAFAC decomposition on the small core tensor $\bG$ \citep{bro1998improving,phan2013candecomp}.  We initialize SPACO  with Algorithm  \ref{alg:initialization}, which combines the above approach with functional PCA \citep{yao2005functional} to work with sparse longitudinal data. Algorithm 
\ref{alg:initialization} consists of the following steps: (1)  perform SVD on $\bX_J$ to get $ \bV_{\perp}$; (2) project $\bX_{J}$ onto each column of $\bV_{\perp}$ and perform
functional PCA to estimate $ \bPhi_{\perp}$; (3) run a ridge-penalized regression of rows of $\bX_{I}$ on $ \bV_{\perp}\otimes \bPhi_{\perp}$, and estimate $ \bU_{\perp}$ and $\bG$ from the regression coefficients

\begin{algorithm}
\caption{Initialization of SPACO}
\label{alg:initialization}
\setcounter{AlgoLine}{0}
Let $\bV_{\perp}$ be the top $K_3$ left singular vectors of $\bX_{J}$ using only the observed columns.

Set $\bY(k) = (\bY_1(k),\ldots,  \bY_T(k)) \in \mathbb{R}^{I \times T}$,
where $\bY_t(k) = \bX_{:,t,:}(\bV_{\perp})_k \in \mathbb{R}^I$.

 Let $\bPhi_{\perp}$ be the top $K_2$ eigenvectors from functional PCA  on the row aggregation of matrices $\bY(k)$ $k=1,\ldots, K_3$. (see Appendix \ref{app:initialization} for details on functional PCA.)

Let $\tilde\bU = \arg\min_{\bU} \{\|\bX_{I} -
\bU(\bV_{\perp}\otimes\bPhi_{\perp})^\top\|_F^2+\delta \|\bU\|_F^2\} \in
\mathbb{R}^{I\times K^2}$ for some small $\delta$. 

Let $\bU_{\perp}$ be the top $K$ left singular eigenvectors of $\tilde
\bU$, and $\tilde{\bG} = \bU^\top_{\perp}\tilde \bU \in \mathbb{R}^{K\times
K^2}$. Let $ \bG\in \mathbb{R}^{K\times K \times K}$ be the estimated core array
from rearranging $\tilde \bG$.

Let $\sum_{k=1}^K \bA_{k}\circledcirc \bB_k \circledcirc \bC_k$ be the
rank-K CP approximation of $\bG$. Stack these as the columns of
$\bA,\bB,\bC \in \mathbb{R}^{K \times K}$, and set $[\bU, \bPhi, \bV] = [\bU_{\perp} \bA,  \bPhi_{\perp}\bB,\bV_{\perp} \bC]$.

For each $k=1,\ldots,K$, rescale the  initializers to satisfy  constraints on $\bV$ and $\bPhi$. 
\end{algorithm}
In a noiseless model with $\delta=0$ and complete temporal observations, one
may replace the functional PCA step of Algorithm \ref{alg:initialization} with
standard PCA. Then $[\bU, \bPhi, \bV]$ becomes a PARAFAC decomposition of
$\frac{1}{1+\delta}\bX$.
\begin{lemma}
\label{lem:initialization}
Suppose $\bX = \sum_{k=1}^K \bU^*_k \circledcirc \bPhi^*_k \circledcirc \bV^*_k$
and is completely observed. Replace $\bPhi_{\perp}$ in Algorithm
\ref{alg:initialization} by the top $K$ eigenvectors of $\bW=\frac{1}{I}
\sum_{k=1}^K \bY(k)^\top\bY(k)$. Then, the outputs $\bU,\bPhi,\bV$ of Algorithm \ref{alg:initialization} satisfy that $\bX = (1+\delta)\sum_{k=1}^K \bU_k \circledcirc \bPhi_k \circledcirc \bV_k$. 
\end{lemma}
Proof of Lemma \ref{lem:initialization} is given in Appendix \ref{app:proof_initialization}.

\subsection{Auto-selection of tuning parameters}
\noindent\textbf{Selection of regularizers $\lambda_1$ and $\lambda_2$:} One way to choose the tuning parameters $\lambda_{1k}$ and $\lambda_{2k}$ is to use cross-validation. However, this
can be computationally expensive even if we tune each parameter sequentially. It is also difficult to determine a good set of candidate values for the parameters
before running SPACO. Instead, we determine the tuning parameters via
nested cross-validation, which has been shown to be empirically useful
\citep{huang2008functional,li2016supervised}. In nested cross-validation, the
parameters are tuned within their corresponding subroutines:

\noindent(1) In the update for $\bPhi_k$, perform column-wise leave-one-out cross-validation to select $\lambda_{1k}$, where we leave out all observations from a single time point. (See Appendix \ref{app:phi_tuning}.)

\noindent(2) In the update for $\bbeta_{:,k}$, perform K-fold cross-validation to select $\lambda_{2k}$.

\noindent\textbf{Rank selection:} We can perform rank selection via cross-validation as suggested in  SupCP \citep{lock2018supervised}.  To reduce the computational cost, we deploy two strategies. (1) Early stopping:  we stop where the cross-validated marginal log-likelihood is no longer decreasing. (2) Warm start and short-run: we initialize the model parameters for each cross-validated model at the global solution and only run a small number of iterations. We found 5 or 10 iterations are usually sufficiently large in practice (the default maximum number of iterations is $10$).

\subsection{Covariate importance} 
In our synthetic experiments in Section \ref{sec:sims}, we observe that the inclusion of $Z$ can result in a better estimate of the subject scores when the true subject scores strongly depend on $Z$. A natural following-up question is if we can ascertain the importance of such covariates when they exist. Here, we consider the construction of approximated p-values from conditional independence\slash marginal independence tests between $\bZ_j$ and $\bU_k$:
\[
H_{0k}^{partial}: Z_j\perp\!\!\!\perp \mu_k|Z_{j^c}, \qquad
H_{0k}^{marginal}: Z_j\perp\!\!\!\perp \mu_k
\]
Both questions are of practical interest. 

\noindent\textbf{Recap on randomization-based hypothesis testing:} Before we proceed to our proposal, we review some basic results for hypothesis testing via randomization. Consider a linear regression problem where $Y=Z\beta+\zeta=f(Z)+\zeta$, with mean-zero noise $\zeta\perp\!\!\!\perp  Z$. Randomization test is a procedure for constructing a valid p-value without assuming the correctness of the linear model of $Y$ on $Z$.

Testing for marginal independent between $Y$ and any covariate $Z_j$ is straightforward.  Let  $t(\bZ_j,\by)$ be a test statistic. Let $T\coloneqq
t(\bZ_j, \by)$ and $T^*\coloneqq t(\bZ_j^*, \by)$ , where
$\bZ_j^*$ is a permutation of $Z_j$. Then, under the null hypothesis $H_0^{marginal}$, $T$ are exchangeable with independent copies of $T^*$. Hence,  $P(T > t^*_{1-\alpha}|\by)\leq \alpha$ for any $\alpha \in (0,1)$ where $t^*_{1-\alpha}$ is the $(1-\alpha)$-percentile of the empirical distribution formed by copies $T^*$  and $\infty$.

Suppose that we have access to the conditional distribution  of $Z_j|Z_{j^c}$. Let $t(\bZ_j,\bZ_{j^c},\by)$ be a test statistic. Let $T\coloneqq t(\bZ_j,\bZ_{j^c},\by)$ and $T^*\coloneqq t(\bZ_j^*,\bZ_{j^c},\by)$, where $\bZ_j^*$ is an  independent copy generated from the conditional distribution $Z_j|Z_{j^c}$ and $\bZ_{j^c}$. Then, under the null hypothesis $H_0^{partial}$, $T$ and $T^*$ have the same law conditional on $\bZ_{j^c}$ and $\by$. So $P(T > t^*_{1-\alpha}|\bZ_{j^c},\by)\leq \alpha$, where $t^*_{1-\alpha}$ is the $(1-\alpha)$-percentile of the conditional distribution of $T^*$ \citep{candes2018panning}.

\noindent\textbf{Oracle randomization test in SPACO:} Let's now go back to SPACO  and consider the ideal setting where   $\bV$, $\bPhi$, $s^2$ and $\sigma^2$ are given. Lemma \ref{lem:testing1} forms the basis of our proposal.
\begin{lemma}
\label{lem:testing1}
Given $\bV, \bPhi, s^2, \sigma^2$, and let $\beta_{\ell}^*$ be the true regression coefficients on the covariates $Z$ for $\mu_{\ell}$, $\ell = 1,\ldots, K$. For any $k=1,\ldots, K$, we define 
{\small
\begin{align}
&\bSigma_i(\delta) =\left(\delta\Lambda_f+(\bV\odot\bPhi)_{\vec i}^\top\Lambda_{\vec i}(\bV\odot\bPhi)_{\vec i}\right)^{-1},\\
&w_i(\delta) =
s_k^2+(1-2\delta)\bSigma_{i}(\delta)_{kk}+(\delta^2-\delta)\bSigma_{i}(\delta)_{k,:}\Lambda_f\Sigma_{i}(\delta)_{:,k},\label{eq:test1}\\
&\tilde z_i(\delta) = \left(1-\delta\frac{\bSigma_i(\delta)_{kk}}{s_k^2}\right)\bz_i, \quad\tilde y_i(\delta) = \bSigma_i(\delta)_{k,:}(\bV\odot \bPhi)^\top\Lambda_{\vec
i}\bX_{I,\vec i}+\delta\sum_{\ell\neq
k}\frac{\bSigma_{i}(\delta)_{k\ell}}{s_{\ell}^2}\bz_i^\top\bbeta_{\ell}.\label{eq:test2}
\end{align}
}
Then, set  $\Delta_i(\delta) = \delta\sum_{\ell\neq k}\frac{\bSigma_{i}(\delta)_{k\ell}}{s_{\ell}^2}\bz_i^\top(\bbeta_{\ell}-\bbeta_\ell^*)$, $\tilde{y}_i(\delta)$ can be expressed as
\begin{equation}
\tilde y_i(\delta) = \tilde \bz_i(\delta)^\top\bbeta_k^*+\Delta_i(\delta)+\xi_i,\quad \xi_i \sim \mathcal{N}(0, w_i(\delta)). \label{eq:test3}
\end{equation}
\end{lemma}
Proof of Lemma \ref{lem:testing1} is given in Appendix \ref{app:proof_testing}. As a result,  when the term $\Delta_i(\delta)=0$, we can apply the traditional randomization tests with response $\tilde y_i(\delta)$ and features $\tilde \bz_i(\delta)$ for subject $i$.  Proposition \ref{prop:randomization_oracle} gives our construction of the observed and randomized test statistics .

\begin{proposition}
\label{prop:randomization_oracle}
Set  $T_{partial}(\delta) = \frac{\sum_{i}\frac{1}{w_i(\delta)}(\tilde y_i(\delta) - \tilde\bz_{i,j^c}^\top(\delta)\bbeta_{j^c,k})\tilde{z}_{ij}(\delta)}{\sum_i \frac{1}{w_i(\delta)} \tilde{z}_{ij}^2(\delta)}$, $T_{marginal}(\delta) = \frac{\sum_{i}\frac{1}{w_i(\delta)}\tilde y_i(\delta)\tilde{z}_{ij}(\delta)}{\sum_i \frac{1}{w_i(\delta)} \tilde{z}_{ij}^2(\delta)}$. Replacing $Z_{j}$ with the  properly randomized $Z^*_{j}$  to create $T_{marginal}^*(\delta)$ and $T^*_{partial}(\delta)$ $(Z^*_j$ is the permutation of $Z_j$ in $T_{marginal}^*(\delta)$ and is independently generated from $Z_j|\bZ_{j^c}$ in $T_{partial}^*(\delta))$.
When   $\Delta(\delta) = 0$, we have
\begin{align}
&T_{marginal}(\delta)|\tilde\by(\delta)\overset{d}{=}T^*_{marginal}(\delta)|\tilde\by(\delta), \qquad \mbox{under $H_{0k}^{marginal}$},\label{eq:stats_marginal_null}\\
&T_{partial}(\delta)|(\tilde\by(\delta), \bZ_{j^c})\overset{d}{=}T^*_{partial}(\delta)|(\tilde\by(\delta), \bZ_{j^c}), \qquad \mbox{under $H_{0k}^{partial}$}.\label{eq:stats_partial_null}
\end{align}
\end{proposition}
In practice, $\bbeta_{\ell}$ for the other factors $\ell\neq k$ are also
estimated from the data. When $\delta\neq 0$, we could have $\Delta_i(\delta)\neq 0$, which renders the randomization test invalid. Thus, we typically set $\delta=0$. For convenience, we drop the $\delta$ argument when $\delta = 0$, e.g.,  $T_{partial}=T_{partial}(\delta)$.  Algorithm \ref{alg:testing_construction} summarizes our proposal.
\begin{algorithm}
\caption{Randomization test for $\bZ_j$}
\label{alg:testing_construction}
\setcounter{AlgoLine}{0}
\For{$k = 1,\ldots,K$}{
\nl Construct responses and features as described in Lemma \ref{lem:testing1},
for $\delta=0$.

\nl Define $\hat \bbeta_k$ by
\[
\hat\bbeta_k =
\arg\min_{\bbeta_k:\bbeta_{k,j}=0}\left\{\sum_{i=1}^I\frac{1}{w_i(0)}(\bz_i^\top
\bbeta_k - \tilde y_i(0))^2+\lambda_{2k}|\bbeta_k|_1\right\}.
\]

\nl Compute the designed test statistic $T$ using $(\bZ_j, \bZ_{j^c},\tilde \by(0), \hat \bbeta_{j^c,k})$.

\nl Compute randomized statistics $T^*_b$ using $(\bZ^{b*}_j, \bZ_{j^c},\tilde
\by(0), \hat \bbeta_{j^c,k})$, where $\bZ^{b*}_j$ for $b=1,\ldots,B$ are (conditionally or marginally) independent copies of $\bZ_j$.

\nl Let $\hat G(.)$ be the empirical estimate of the CDF of $T$ under $H_0$
using $\{T_1^*, \ldots, T_B^*\}$, and return the two-sided p-value
$p=[1-\hat G(|T|)]+\hat G(-|T|)$.}
\end{algorithm}
\begin{remark}
For $\delta>0$, the test based on $\tilde{y}_i(\delta)$,
$w(\delta)$, and $\tilde{\bz}_i(\delta)$ trades off robustness against
estimation errors in $Z^\top(\bbeta_{\ell}-\bbeta^*_{\ell})$ for possibly
increased power. To see this, suppose that $\bbeta_{\ell}=\bbeta_{\ell}^*$ for $\ell\neq k$ and
$Z_j\sim \mathcal{N}(0,1)$. The signal to noise ratio with $\delta\in \{0,1\}$
can be calculated as
\begin{align*}
{\rm SNR}(0)&=\frac{\bE(\bz_i^\top\bbeta_k^*)^2}{\frac{1}{I}\sum_i w_i(0)}
=\frac{\|\bbeta_k^*\|_2^2}{s_k^2+\frac{1}{I}\sum_i \bSigma_i(0)_{kk}},
{\rm SNR}(1) =\frac{\frac{1}{I}\sum_i
\bE(\tilde{\bz}_i(1)^\top\bbeta_k^*)^2}{\frac{1}{I}\sum_i w_i(1)}=\frac{\|\bbeta_k^*\|_2^2}{s_k^2}.
\end{align*}
Thus, the signal-to-noise ratio is higher with $\delta = 1$ if we have access to the true $\bbeta_\ell^*$. 
\end{remark}
The conditional randomization test requires generating $Z_j^*$ from the conditional distribution $Z_j|Z_{j^c}$.  We estimate the conditional distribution of $Z_j^*$ via proper exponential family distribution . More details on the generations of $\bZ_j^*$ are provided in Appendix \ref{app:testing_Zgenerate}.

\noindent\textbf{Approximated p-value construction with estimated model paramters:} The true model parameters for $\bV$, $\bPhi$, $s^2$ and $\sigma^2$ are of course unknown.  We will substitute their empirical estimates in the above procedure. However, the substitutes from a full SPACO fit may suffer from fitting towards the empirical noise. To reduce the influence of over-fitting,  we use $\bV, \bPhi$ from cross-validation as described for performing the rank selection. That is,  for $i\in \mathcal{V}_m$ where $\mathcal{V}_m$ is the index set for fold $m$ in cross-validation, we construct  $\tilde y_i(0) = \Sigma_i(0)_{k,:}(\bV^{-m}\odot \bPhi^{-m})^\top\Lambda_{\vec i}\bX_{I,\vec i}$, where  $\bV^{-m}$, $\bPhi^{-m}$ are estimates using fold other than $\mathcal{V}_m$. $\Sigma_i(0)$ is also estimated using  $\bV^{-m}$, $\bPhi^{-m}$ and cross-validated prior covariance $\Lambda_f^{-m}$. We refer it as cross-fit, where we estimate some model parameters using data from other folds and update the quantity of interest using them and the current fold. Since each fold is initialized at the global solution, this further encourages the comparability of the constructed $\tilde y$ and $\tilde z$ from different folds. We observe that the cross-fit offers better Type I error control than the naive plug-in of the full estimates.
\section{Numerical studies with synthetic data}
\label{sec:sims}
In this section, we evaluate SPACO with synthetic Gaussian data. We fix the variance at 1 for the noise $\epsilon$ and the number of true ranks $K=3$. We consider simulated data $(\bX, \bZ)$ with dimensions  $(I, T, J, q)=(100, 30, 10, 100)$ and $(100, 30, 500, 100)$, with the observed rate ($=1-$missing rate)  along the time dimension $r\in \{100\%, 50\%, 10\%\}$ and observed time stamps chosen randomly for each subject. Given the dimension and the observed rate, we  generate $v_{jk}\overset{i.i.d}{\sim} \mathcal{N}(0, \frac{1}{J})$ and $z_{i\ell}\overset{i.i.d}{\sim} \mathcal{N}(0,1)$ for $i=1,\ldots, I$, $j=1,\ldots, J$ and $\ell = 1,\ldots, q$. Then, (1)we set $\phi_1(t) = \theta_1$, $\phi_2(t) = \theta_2\sqrt{1-\left(\frac{t}{T}\right)^2}$, $\phi_3(t)= \theta_3\cos(4\pi\frac{t}{T})$ with random parameters $\theta_1, \theta_2, \theta_3\sim c_1 \cdot \mathcal{N}(0, \frac{\log J+\log T}{nT\gamma})$ for $c_1\in \{1, 3, 5\}$, and (2) we set $\beta_{\ell,k} \sim c_2 \cdot \mathcal{N}(0, \frac{\log q}{I})$ for $c_2\in \{0, 3, 10\}$ for the first $\ell = 1, 2, 3$, and set $\beta_{\ell k}=0$ otherwise.  Each $\bU_k$ is standardized to be mean 0 and variance 1 after generation. This generates 54 different simulation setups in total.
\subsection{Reconstruction quality evaluation}
\label{subsec:sim1}
\begin{figure}[H]
\caption{Reconstruction evaluation by the correlations between the estimates and the true signal tensor. In each subplot, x axis label indicates different $J$ and observing rate,  the y axis is the achieved correlation, and  the box colors represent different methods. The  corresponding subplot column/row name represents the signal-to-noise ratio SNR1\slash SNR2.}
\label{fig:construction_signalmat}
\includegraphics[width  = 1.0\textwidth]{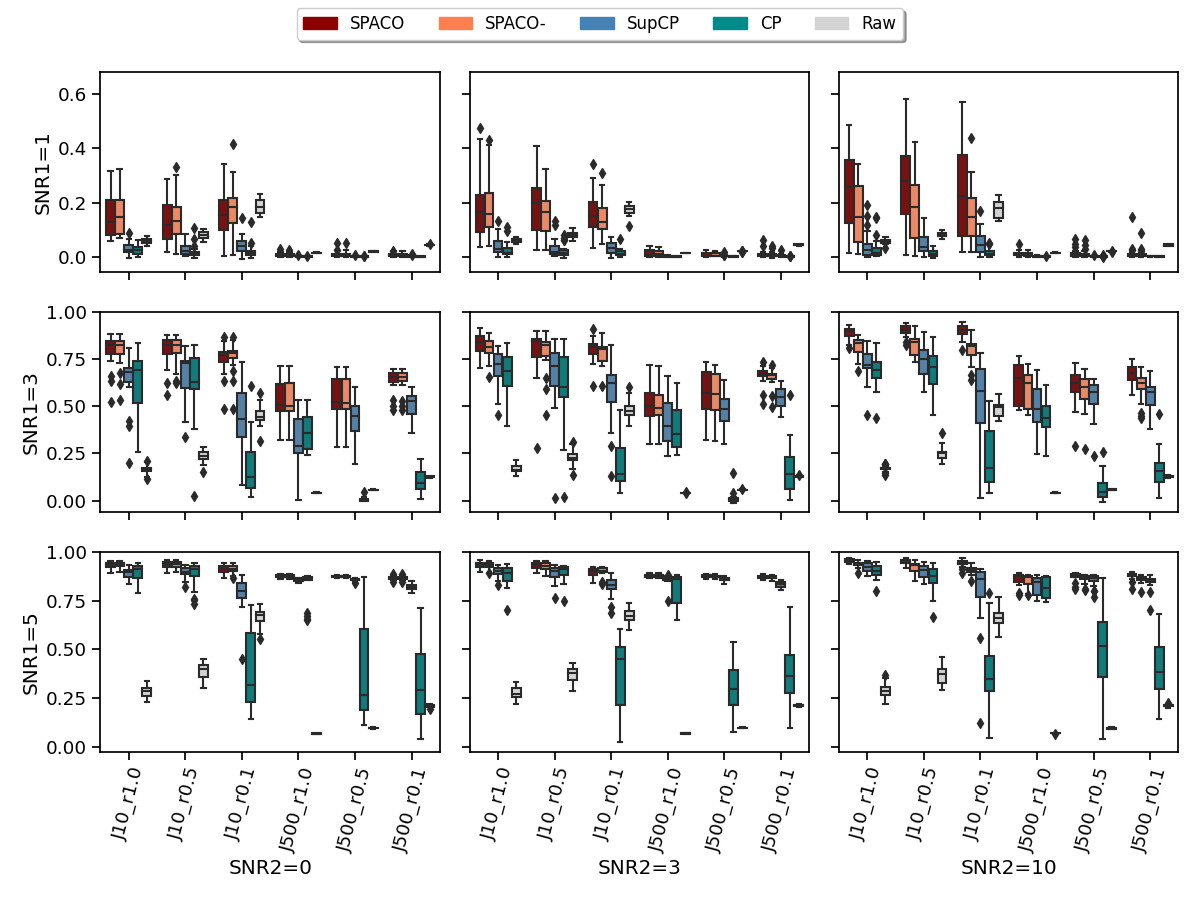}
\end{figure}

\begin{figure}[H]
\caption{Comparison of SPACO and SPACO- for reconstructing $U$ at $J=10$, SNR2$=10$. In each subplot, x axis label indicates different component and observing rate,  the y axis is the achieved $(1-R^2)$, and  the box colors represent different methods. The  corresponding subplot column/row name represents the signal-to-noise ratio SNR1\slash component.}
\label{fig:construction_U}
\includegraphics[width  = 1.0\textwidth, height = 0.6\textwidth]{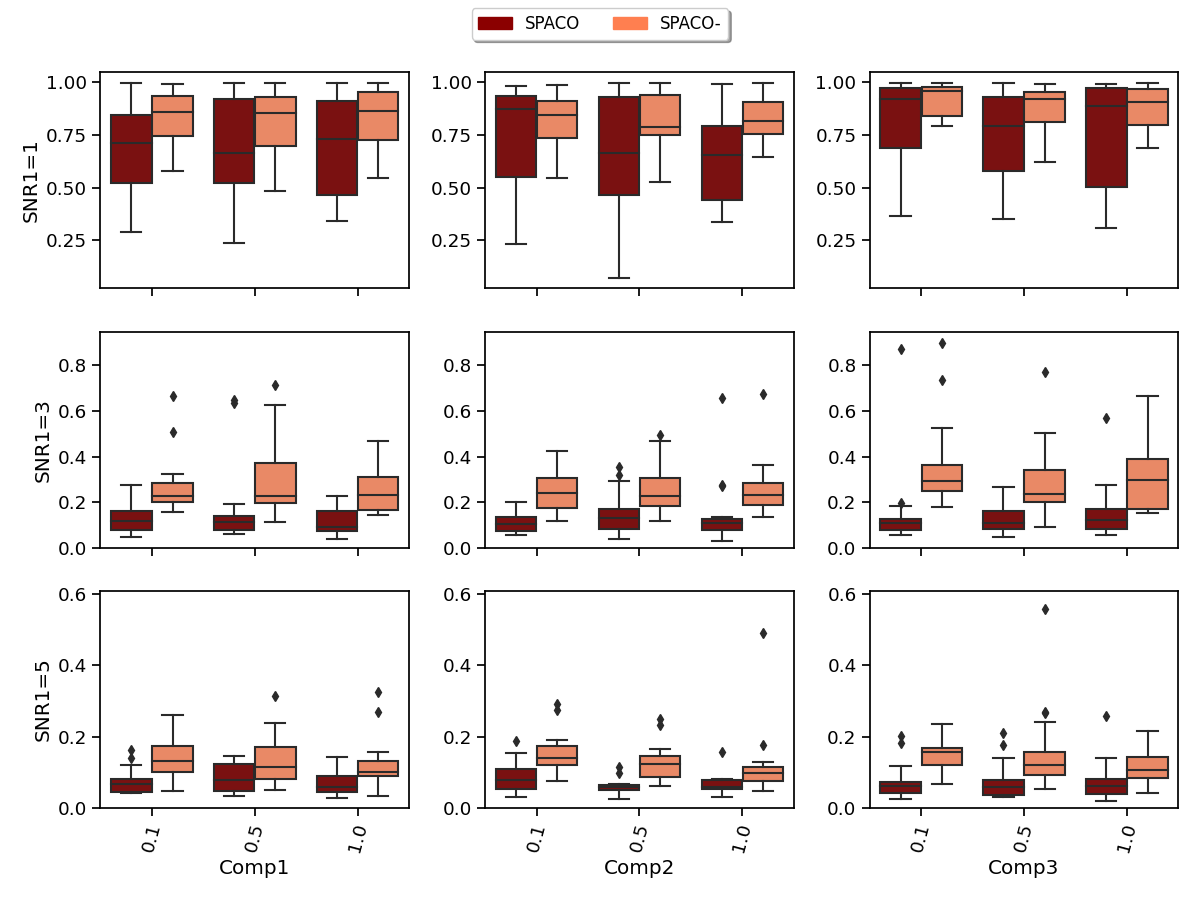}
\end{figure}

\begin{figure}
\caption{Comparison of SPACO and SupCP for reconstructing $\Phi$ at $J=10$. In each subplot, x axis label indicates different component and observing rate,  the y axis is the achieved $(1-R^2)$, and  the box colors represent different methods. The  corresponding subplot column/row name represents the signal-to-noise ratio SNR1\slash SNR2.}
\label{fig:construction_phi}
\includegraphics[width  = 1.0\textwidth]{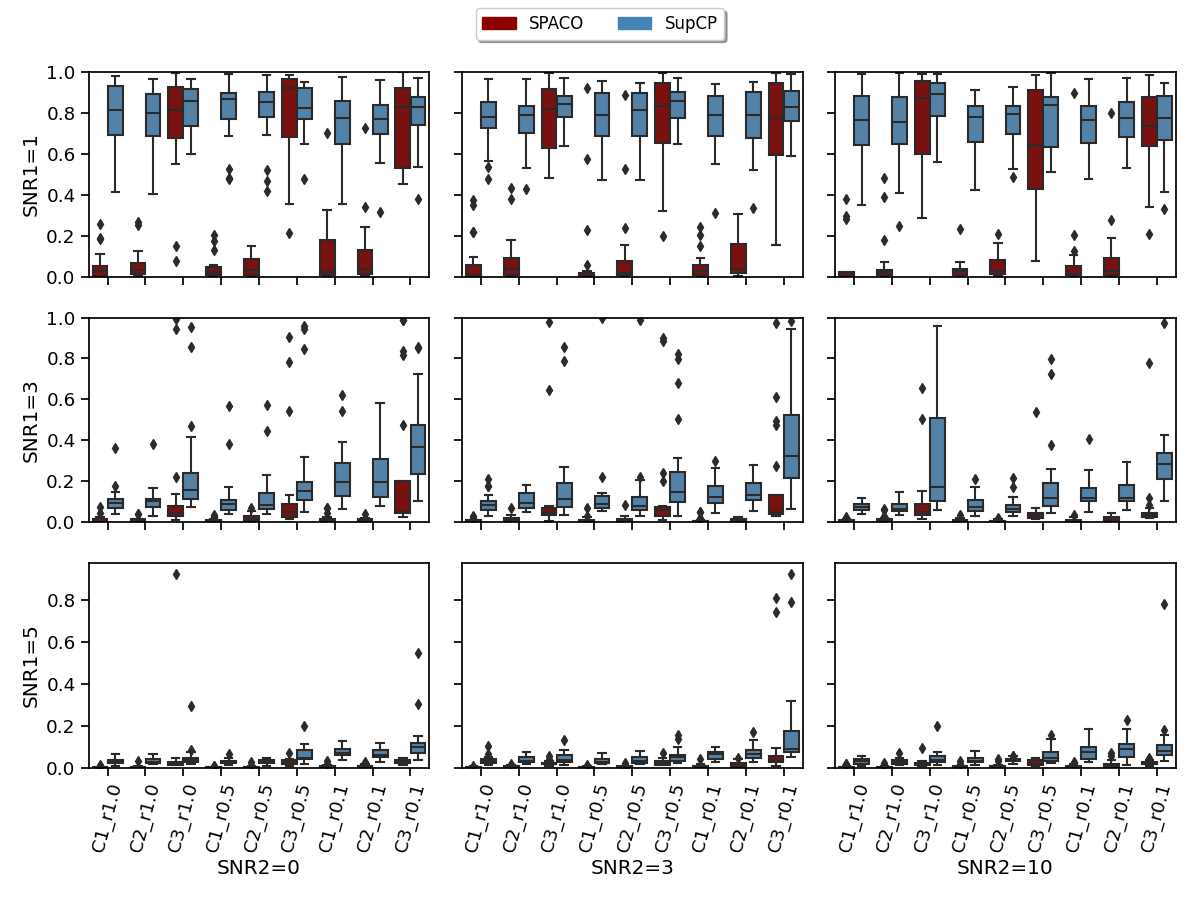}
\end{figure}

We compare SPACO, SPACO-, plain CP from python package {\it tensorly}, and a proxy to SupCP by setting  $\lambda_{1k} = \lambda_{2k} = 10^{-2}$ as small fixed values in SPACO (the additional small penalties improve numerical stability to deal with large $q$ and high missing rate). Although this is not exactly SupCP, they are very similar. We refer to this special case of SPACO as SupCP and include it to evaluate the gain from smoothness regularization on the time trajectory. Note that we use the proposed initialization in SPACO, SPACO- and SupCP, which provide better and more stable results when the missing rate is high. A comparison between SupCP with random and proposed initialization is given in Appendix \ref{app:numerical_init}. We use the true rank in all four methods in our estimation. Figure  \ref{fig:construction_signalmat}   shows the achieved correlation between the reconstructed tensors and the underlying signal tensors across different setups and 20 random repetitions. SNR1 represents $c_1$, SNR2 represents $c_2$, and each sub-plot represents different signal-to-noise ratios SNR1 and SNR2, as indicated by its row and column names. The y and x axes indicate the achieved correlation and different combinations of $J$ and observed rate, respectively. For example, x-axis label $J10\_r0.1$ means the feature dimension is 10, and 10\% of entries are observed. The ``Raw" method indicates the results by correlating the true signal and the empirical observation on only the observed entries. It is not directly comparable to others with missing data, but we include it here to show the signal level of different simulation setups. We also compare the reconstruction quality on missing entries in Appendix \ref{app:numerical_reconstruct}.
SPACO outperforms SPACO- when the subject score $U$ depends strongly on $Z$, which could result from a more accurate estimation of the subject scores. To confirm this, we evaluate the estimation quality of $U$ at $J = 10$ and SNR2$=10$ and measure the estimation quality by $R^2$ (regressing the true subject scores on the estimated ones). In Figure \ref{fig:construction_U}, we shows the achieved $(1-R^2)$ for SPACO and SPACO- (smaller is better), where  x-axis label represents  the observing rate and column names represent the component, e.g., Comp1$\rightarrow U_1$. 
SPACO \slash SPACO- are both top performers for our smooth tensor decomposition problem and achieve significantly better performance than CP and SupCP when the signal is weak and when the missing rate is high by utilizing the smoothness of the time trajectory. To see this, we compared the estimation quality of the time trajectories using SPACO and SupCP. In Figure \ref{fig:construction_phi}, we shows the achieved $(1-R^2)$ for SPACO and SupCP at $J=10$. In the x-axis label represents different trajectory and observed rate, e.g., ($C1_r1.0\rightarrow$ estimation of $\phi_1(t)$ at observing rate $r=1.0$). When the signal is weak  (SNR1=1), SPACO could approximate the constant trajectory component ($C1$) and start to estimate other trajectories successfully as the signal increases. SPACO achieves significantly better estimation of the true underlying trajectories than SupCP for various signal-to-noise ratios.
\subsection{Variable importance and hypothesis testing}
We investigate the approximated p-values based on cross-fit for testing the partial and marginal associations of $\bZ$ with $\deta$ under the same simulation set-ups. Since our variables in $\bZ$ are generated independently, the two null hypotheses coincide in this setting. However, the two tests have different powers given the same p-value cut-off because the test statistics differ. 

\begin{figure}[H]
\begin{center}
\caption{Achieved power at observing rate $r = 0.5$. In each subplot, x axis label indicates different combination of feature dimension $J$ and targeted level $\alpha\in \{0.01, 0.05\}$,  the y axis is the achieved power. Different bar colors represent different tests (partial or marginal). }
\label{fig:power_main}
\includegraphics[width = .9\textwidth, height = .7\textwidth]{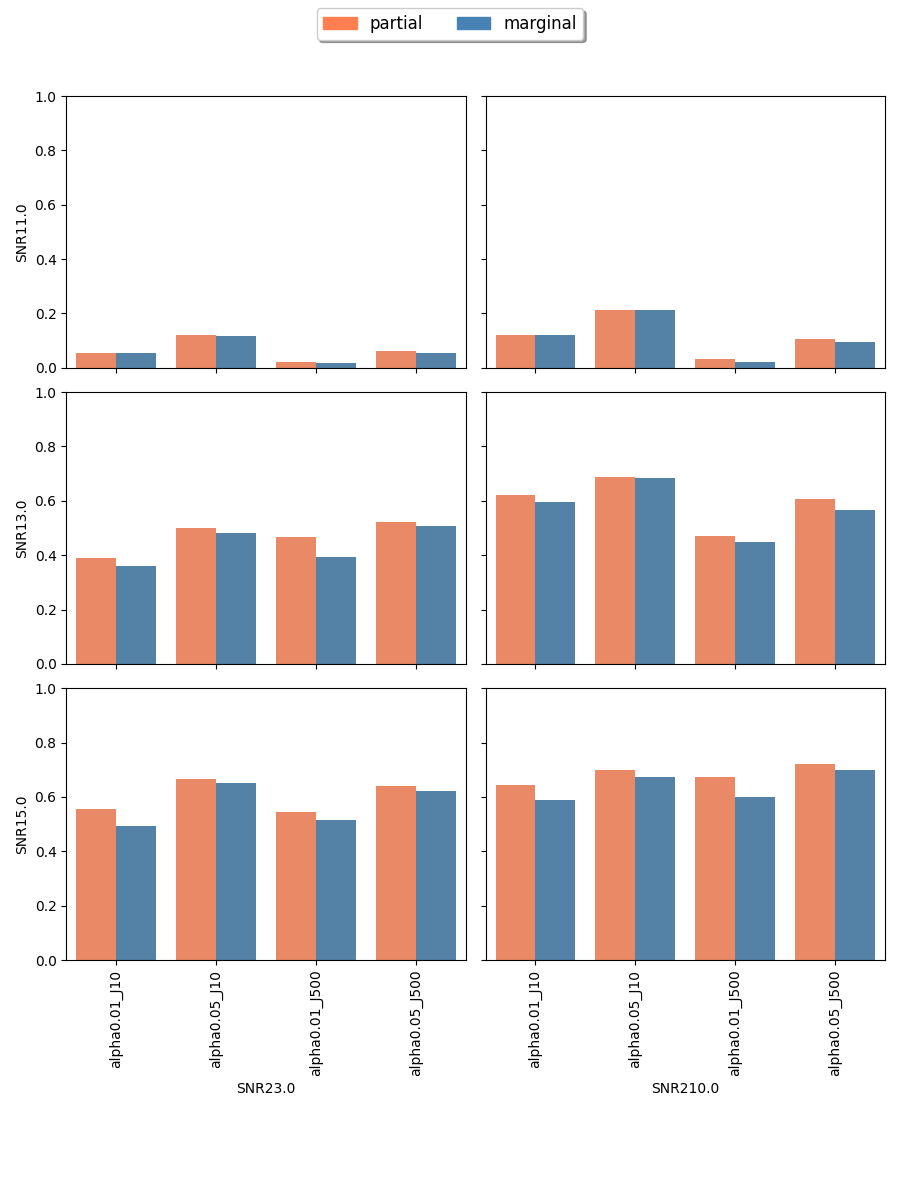}
\end{center}
\end{figure}

The proposed randomization tests for SPACO achieve reasonable Type I error controls. Cross-fit is important for a good  Type I error control: In Appendix \ref{app:cross_naive_pval}, we present qq-plots comparing p-values using cross-fitted $\bV$ and $\bPhi$ and the naive construction. We observe noticeable deviations from the uniform distribution for the later construction when the signal-to-noise ratio is low. Fig.\ref{fig:typeI_main} and Fig.\ref{fig:power_main} show the the achieved Type I error and power with p-value cut-offs at $\alpha =0.01, 0.05$ and with observed rate $r = 0.5$. The type I errors are also well controlled for $r\in\{1.0,0.1\}$ (Appendix \ref{app:numerical_typeI}).

\begin{figure}[H]
\begin{center}
\caption{Achieved type I errors at observing rate $r = 0.5$. In each subplot, x axis label indicates different combination of feature dimension $J$ and targeted level $\alpha\in \{0.01, 0.05\}$,  the y axis is the achieved type I errors. Different bar colors represent different tests (partial or marginal). The two dashed horizontal lines indicate levels 0.01 and 0.05.}
\label{fig:typeI_main}
\includegraphics[width  = .9\textwidth, height = .67\textwidth]{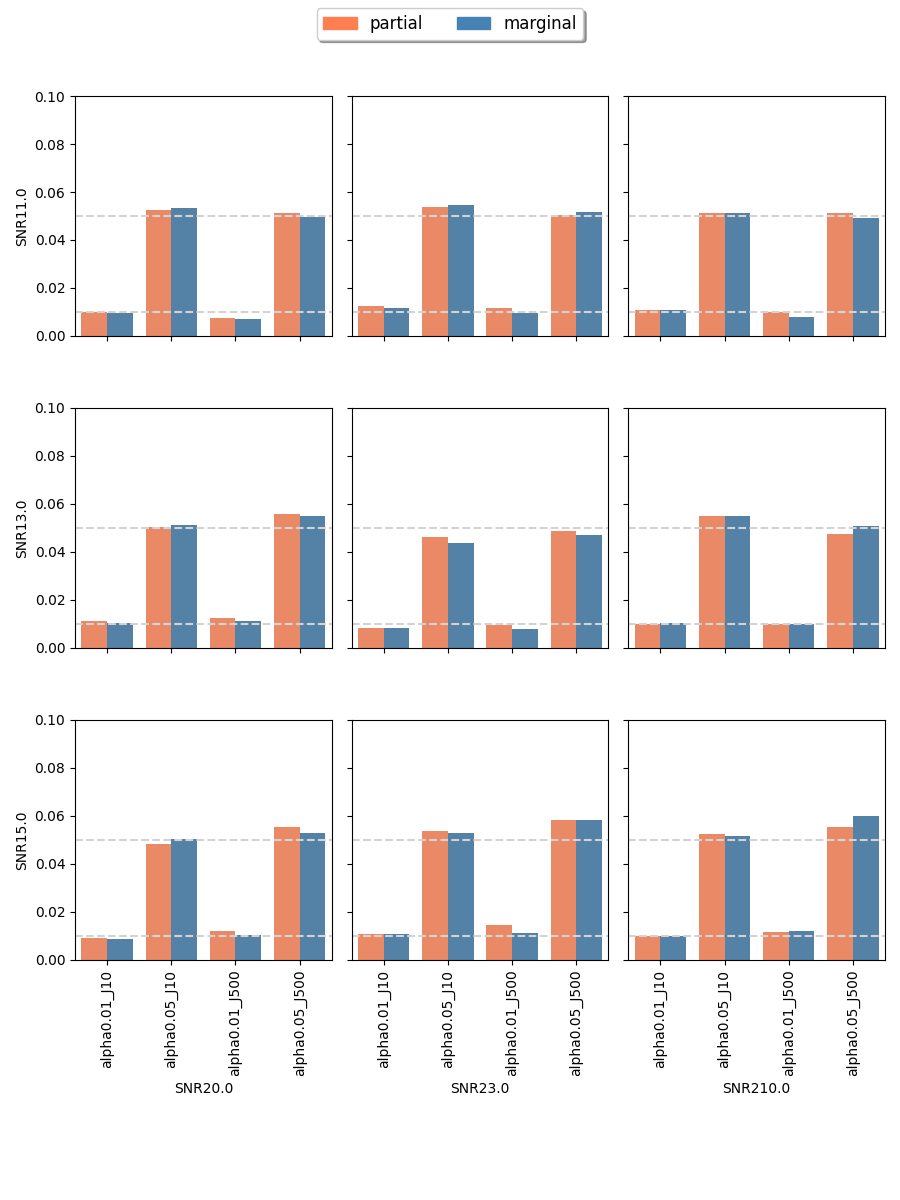}
\end{center}
\end{figure}

\section{Case study}
\label{sec:case_study}
We now apply SPACO to a longitudinal immunological data set on COVID 19 from the IMPACT study \citep{lucas2020longitudinal}. Initially, the data contained 180 samples on 135 immunological features, collected from 98 participants with COVID-19 infection. We filter out features with more than $20\%$ missingness among collected samples and imputed the missing observations using MOFA \citep{argelaguet2018multi} for the remaining features. This leaves us with a complete matrix with 111 features and 180 samples, which is organized into a tensor of size $(I, T, J) = (98, 35, 111)$ where $T$ is the number of unique DFSO (days from symptom onsite) in this data set. This is a sparsely observed tensor, and the average observing rate is 1.84 along the time dimension. Apart from the immunological data, the data set also provides non-immunological covariates. We use eight static risk factors as our covariate $\bZ$ (COVID\_risk1 - COVID\_risk5, age, sex, BMI), and four symptom measures as additional responses (ICU, Clinical score, LengthofStay, Coagulopathy). 
\begin{figure}[H]
\caption{Left  panel shows examples of time trajectories of four features
for different subjects (horizontal time axis is the DFSO ) ; right panel shows plots of
observed feature values against estimated ones from SPACO. Each line/dot represents
observations from a single subject. Subjects in ICU are colored red. }
\label{fig:IMPACT_overview}
\begin{flushleft}
\includegraphics[width  = 1.0\textwidth]{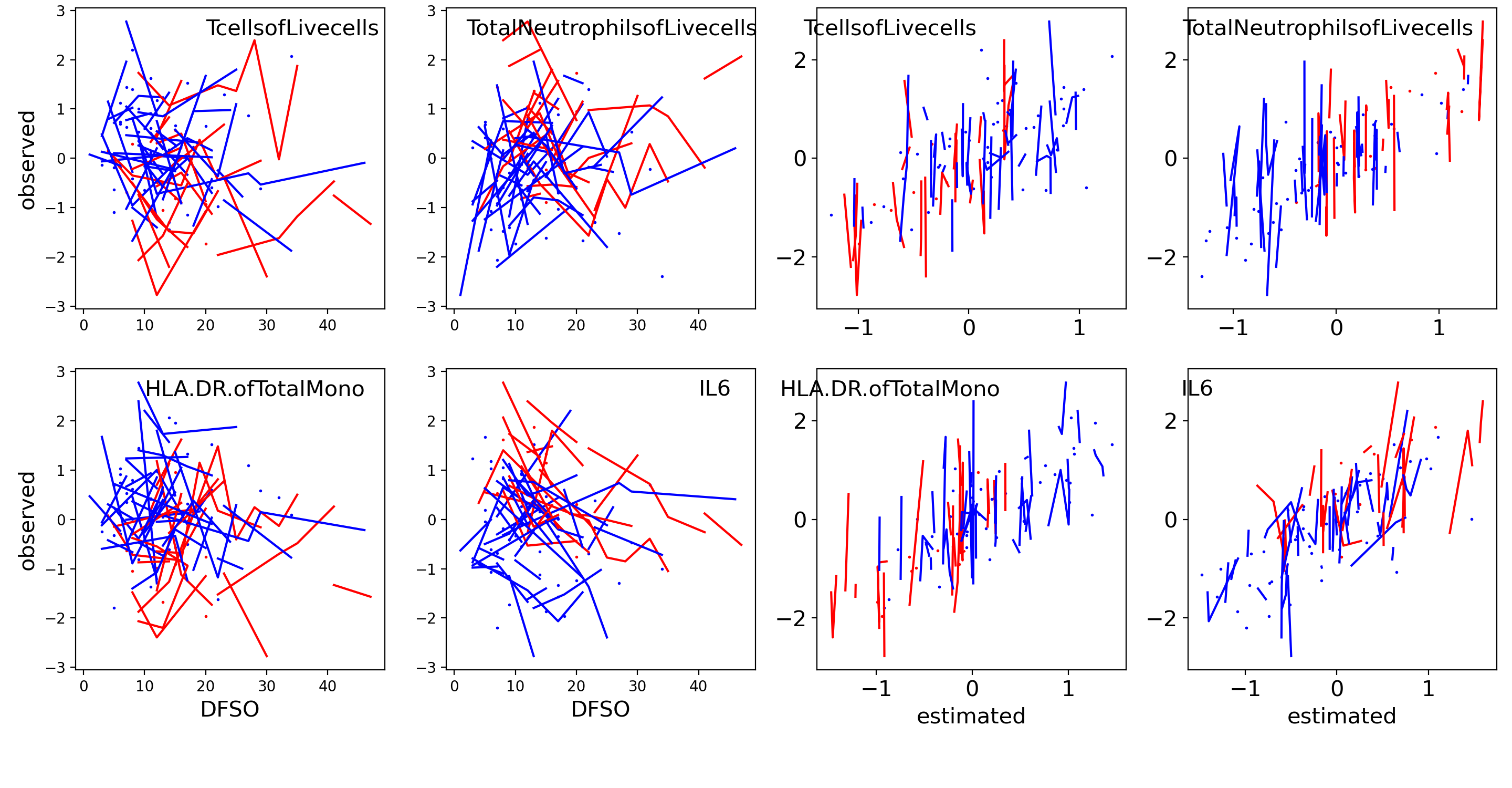}
\end{flushleft}
\end{figure}
We run SPACO with processed $\bX$, $\bZ$, and model rank $K=4$, selected from finding the first maxima for the five-fold cross-validated marginal log-likelihood. Fig.\ref{fig:IMPACT_overview} shows example plots of raw observations against our reconstructions across all samples. We see a positive correlation between these observed and reconstructed values.  

Combining static covariates $Z$ with the longitudinal measurements can sometimes improve the estimation quality of subject scores compared to SPACO-, as we have illustrated in our synthetic data examples. We can not possibly get the true subject scores in the real data set. However, we could still compare the estimated subject scores are clinically relevant by comparing them to the responses. Here, we also run SPACO- with $K=4$ and examine if the resulting subject scores from SPACO associate better with the responses. Table  \ref{tab:association_response} shows the spearman correlations and their p-values when comparing the estimated subject scores and each response. The top three components C1, C2, C3 are significantly associated with the responses. Among them, $C_2$ uniquely captures variability for the length of stay (in hospital), and only $C3$ heavily depends on $\bZ$ (the estimated coefficients are $0$ for both $C1$ and $C2$). C3 from SPACO achieved a better association with the clinical outcomes than SPACO.
\begin{table}
\caption{Association between estimated subject scores and four response variables. C1-C4 represent the four components in SPACO\slash SPACO-. C*cor represents the achieved spearman correlation and C*pval is the p-value for the corresponding correlation test. }
\label{tab:association_response}
\begin{adjustbox}{width = \textwidth}
\begin{tabular}{llrlrlrlrl}
\toprule
       &              &  C1cor &   C1pval &  C2cor &   C2pval &  C3cor &   C3pval &  C4cor &   C4pval \\
\midrule
SPACO & ICU &  -0.20 &  4.9E-02 &   0.37 &  1.7E-04 &   0.28 &  5.4E-03 &   0.00 &  9.8E-01 \\
       & Clinicalscore &  -0.32 &  1.2E-03 &   0.42 &  1.3E-05 &   0.44 &  6.4E-06 &   0.07 &  4.8E-01 \\
       & LengthofStay &  -0.10 &  3.4E-01 &   0.28 &  5.2E-03 &   0.14 &  1.8E-01 &   0.11 &  2.9E-01 \\
       & Coagulopathy &  -0.03 &  7.7E-01 &   0.22 &  3.1E-02 &   0.21 &  3.5E-02 &   0.03 &  7.9E-01 \\
SPACO- & ICU &  -0.20 &  5.1E-02 &   0.38 &  1.4E-04 &   0.25 &  1.5E-02 &  -0.00 &  9.6E-01 \\
       & Clinicalscore &  -0.32 &  1.3E-03 &   0.43 &  1.0E-05 &   0.40 &  4.5E-05 &   0.07 &  4.9E-01 \\
       & LengthofStay &  -0.09 &  3.9E-01 &   0.29 &  4.4E-03 &   0.10 &  3.5E-01 &   0.10 &  3.3E-01 \\
       & Coagulopathy &  -0.03 &  8.0E-01 &   0.22 &  3.0E-02 &   0.19 &  5.9E-02 &   0.02 &  8.2E-01 \\
\bottomrule
\end{tabular}
\end{adjustbox}
\end{table}
Using the randomization test, we can also test for the contribution of each $Z_j$ to C3. Table \ref{tab:association_test} shows the p-value and adjusted p-value from the partial dependence test and marginal dependence test, with the number of randomization $B = 2000$. The top associate is COVIDRISK\_3 (hypertension) with a p-value around 0.01 (adjusted p-value around 0.1). BMI is also weakly associated with C3, with p values $< 0.05$ for both tests. In this data set, we observe that the immunological measurements contain information strongly related to clinical responses of interest. Although some risk factors like hypertension and BMI contains relatively weaker signal, they may still improve the estimated subject scores like in the case of C3.
\begin{table}
\caption{Results from randomization test for component 3 (C3). The column nonzero counts the number of 1 for binary covariate COVIDRISK\_1 - COVIDRISK\_5 and sex (1 = Male, 0 = Female). The adjusted pvalues (pval*) are based on BH correction. }
\label{tab:association_test}
\begin{adjustbox}{width = \textwidth}
\begin{tabular}{lrrrrr}
\toprule
{} &  pval(partial) &  pval(marginal) &  padj(partial) &  padj(marginal) &  nonzero \\
\midrule
COVIDRISK\_1 &          0.877 &           0.905 &          1.000 &           0.905 &       7 \\
COVIDRISK\_2 &          0.263 &           0.063 &          0.701 &           0.169 &      24 \\
COVIDRISK\_3 &          0.013 &           0.011 &          0.101 &           0.088 &      50 \\
COVIDRISK\_4 &          0.563 &           0.712 &          1.000 &           1.000 &      23 \\
COVIDRISK\_5 &          0.919 &           0.885 &          0.919 &           1.000 &       5 \\
Age         &          0.584 &           0.423 &          0.935 &           0.846 &       -- \\
sex         &          0.715 &           0.727 &          0.953 &           0.970 &      47 \\
BMI         &          0.029 &           0.041 &          0.117 &           0.166 &       -- \\
\bottomrule
\end{tabular}
\end{adjustbox}
\end{table}

\section{Discussion}
We propose  SPACO to model sparse multivariate longitudinal data and auxiliary covariates jointly. The smoothness regularization may lead to a dramatic improvement in the estimation quality with high missing rates, and access to informative auxiliary covariates could improve the estimation of subject scores. We applied the proposed pipeline to COVID-19 data sets. Even though both data sets are highly sparse and of small size, SPACO identifies components whose subject scores are strongly associated with clinical outcomes of interest and identify static covariates that might contribute to the severe symptoms.   A future direction is to extend SPACO to model multi-omics data. Different omics can be measured at different times and have distinct data types or scales. These motivate a tailored model's more careful design rather than a naive approach of blindly pooling all omics data together.
\begin{center}
{\large\bf SUPPLEMENTARY MATERIAL}
\end{center}

\begin{description}
\item[Python package for SPACO:] \url{https://github.com/LeyingGuan/SPACO}

\item[Python pipeline for reproducing results in the manuscript:]  \url{https://github.com/LeyingGuan/SPACOexperments}
\end{description}

\appendix
\section{Additional Algorithmic Details}
\label{app:algorithm}
\subsection{Updating rules in Algorithm \ref{alg:basicAlg}}
\label{app:alg_basic}
Lemma \ref{lem:update} provides exact parameter update rules used in  Algorithm \ref{alg:basicAlg}. We also define $\langle . \rangle$ as the expectation of some quantity with respect to the posterior distribution of $\bU$. Let $O\in \{0,1\}^{I\times T}$ with $O_{it}=1$ if time $t$ is observed for subject $i$ and $O_{it} = 0$ otherwise.
\begin{lemma}
\label{lem:update}
The parameter update steps in Algorithm \ref{alg:basicAlg} take the following
forms:
\begin{itemize}
\item  In line 4, $\bbeta_k = \arg\max_{\bbeta_k}\{L(\bX|\Theta)+\lambda_{2k}|\bbeta_{.k}|\}=\arg\min_{\bbeta_{k}}\frac{1}{2}\|\tilde y - \tilde \bZ\bbeta_k\|_2^2+\lambda_{2k}|\bbeta_k|$, where
\begin{align}
\tilde \bz_i &=
\sqrt{\frac{1}{s_k^2}\frac{s_k^2-(\Sigma_i)_{k,k}}{s_k^2}}\bz_i,\label{eq:ziupdate}\\
\tilde y_i&= \sqrt{\frac{1}{s_k^2}\frac{s_k^2}{s_k^2 -
(\Sigma_i)_{k,k}}}\left((\Sigma_i)_{k,:}(\bV\odot \bPhi)^\top_{\vec i}\Lambda_{\vec
i}\bX_{I,\vec i}+\sum_{l\neq k}(\Sigma_i)_{k,l}/s_{l}^2\left(\bz_i^\top \bbeta_l\right)\right).
\label{eq:yiupdate}
\end{align}
Here $(\Sigma_i)_{k,:}$ is the $k^\text{th}$ row of $\Sigma_i$, and
$(\Sigma_i)_{k,\bar{k}}$ is this row with $k^\text{th}$ entry removed; $\Lambda_{f,\bar{k}}$ is the sub matrix of $\Lambda_{f}$ with the $k^{th}$ column and row removed.
\item  Set $Q_1 = \langle\bU\rangle\odot\bV\in \real^{IJ\times K}$,   $Q_2\in \real^{K\times J}$ and $Q_3\in \real^{K\times K\times J}$ with $Q_{2kj} = \frac{1}{\sigma_j^2}\bX_{Jj\vec j}^\top Q_{1\vec j k}$ and $Q_{3k_1k_2 j}= \frac{1}{\sigma_j^2}\sum_{t}\phi_{tk_1}\phi_{tk_2}\sum_{O_{it}=1}\left(\Sigma_{ik_1 k_2}+\langle u_{ik_1}\rangle\langle u_{ik_2}\rangle \right)$. In line 5, the update of $\bV_k$ considers the following problem:
\[
\bV_k \leftarrow \arg\min_{\bV_k} [\sum_{j=1}^J\frac{1}{2} \left(a_{jk} v_{jk}^2 -2b_{jk} v_{jk}\right)+\nu \|\bV_k\|_2^2],
\]
where $a_{jk} = \frac{1}{\sigma_j^2}\|Q_{1,\vec j, k}\|_2^2+Q_{3kk,j}$, $b_{jk} =Q_{2kj} - \sum_{k'\neq k}Q_{3kk',j}v_{jk'}$ and $\nu$ is the largest value such that the solution satisfies $\|\bV_k\|_2^2=1$.
\item   Set  $Q_1=\langle\bU\rangle\odot\bPhi\in \real^{IT\times K}$,  $Q_2\in \real^{K\times T}$ and $Q_3\in \real^{K\times K\times T}$ with $Q_{2kt} = \bX_{Tt\vec t}^\top\diag\{\frac{1}{\sigma_{\vec t}^2}\} Q_{1\vec t k}$ and $Q_{3k_1k_2 t}=\sum_{j}\frac{1}{\sigma_j^2}v_{jk_1}v_{jk_2}\sum_{i:O_{it}=1}\left(\Sigma_{ik_1 k_2}+\langle u_{ik_1}\rangle\langle u_{ik_2}\rangle \right)$. In line 5, the update of $\bV_k$ considers the following problem:
\[
\bPhi_k \leftarrow \arg\min_{\bPhi_k} [\sum_{t=1}^T\frac{1}{2} \left(a_{jk} v_{jk}^2 -2b_{jk} v_{jk}\right)+\lambda_{1k}\bphi_k^\top\Omega\bphi_k+\nu \|\bPhi_k\|_2^2],
\]
where  $a_{tk} =\left( \|\diag\{\frac{1}{\sigma_{\vec t}}\}Q_{1,\vec t, k}\|_2^2+Q_{3kk,t}\right)$, $b_{jk} =\left(Q_{2kt} - \sum_{k'\neq k}Q_{3kk',t}\phi_{tk'}\right)$ and $\nu$ is the largest value such that the solution satisfies $\|\bPhi_k\|_2^2=T$.
\item In lines 14 and 15,
\[
s^2_k \leftarrow \frac{1}{I}\sum_i
\left[(\mu_{ik}-\bz_i^\top\bbeta_k)^2+(\bSigma_i)_{kk}\right],
\]
and
\[\sigma_j^2\leftarrow \frac{1}{|\vec j|}
\left[(X_{J,\vec j}-(\bPhi\odot \bmu)_{\vec j}
\bv_{j})^\top(X_{J,\vec j}-(\bPhi\odot \bmu)_{\vec j}
\bv_{j})+\bv_j^\top\left(\sum_{t}\left(\sum_{i:O_{it}=1}\bSigma_{i}\right)\cdot
(\bphi_{t}\bphi_t^\top)\right)\bv_j\right].
\]
\end{itemize}
\end{lemma}  
Derivation of Lemma \ref{lem:update} is given in Appendix \ref{app:proof_update}.
\subsection{Functional PCA for initializations}
\label{app:initialization}
In \cite{yao2005functional}, the authors suggest a  functional PCA analysis by performing eigenvalue decomposition of the smoothed matrix fitted with a local linear surface smoother. Here, we apply the suggested estimation approach on the total  product matrix: Let $\hat \bW_{i,s,t} = \sum_{k}\bY_{is}(k)\bY_{it}(k)$ be the empirical estimate of the total product matrix for  subject $i$,  we first estimate $\bE[\hat W_{i,s,t}]$ via local surface smoother and then perform PCA on the estimated $\bE[\hat W_{i,s,t}]$:
\begin{itemize}
\item  To fit a local linear surface smoother for the off-diagonal element of $\bW_{s_0,t_0}$, we consider the following problem:
\[
\min \sum_i\sum_{O_{it}O_{is}=1,s\neq t} \kappa(\frac{s-s_0}{h_G}, \frac{t - t_0}{h_G})(\hat \bW_{i,s,t}-g((s_0, t_0),(s,t),\beta))^2,
\]
with $g((s_0, t_0),(s,t),\beta) = \beta_0 + \beta_1(s-s_0)+\beta_2(t-t_0)$, and $\kappa:\real^2\mapsto\real$ is a standard two-dimensional Gaussian kernel. 
\item For the diagonal element, we estimate it by local linear regression: for each $t_0$:
\[
\min \sum_i\sum_{O_{it}=1} \kappa_1(\frac{t-t_0}{h_G})(\hat \bW_{i,t,t}-g(t_0,t,\beta))^2.
\]
where $g(t_0, t,\beta) = \beta_0+\beta_1(t-t_0)$.
\end{itemize}
By default, we let $h_G = \frac{T}{\sqrt{1+\sum_{s\neq t}\mathbbm{1}\{\sum_{i}O_{is}O_{it}}>0\}}$.
\subsection{Parameter tuning on $\lambda_{1k}$}
\label{app:phi_tuning}
In this section, we provide more details on  the leave-one-time-out cross-validation for tuning $\lambda_{1k}$, $\forall \; k = 1,\ldots, K$. From eq.\ (\ref{eq:expected_Phi}), the expected penalized deviance loss can be written as (keeping only terms relevant to $\bPhi$): 
\begin{align*}
L&= \sum^\top_{t=1}\left\{\bphi_t^\top\left[(\bV\bigodot \bmu)_{\vec t .}^\top\Lambda^{-1}_{\vec t}(\bV\bigodot\bmu)_{\vec t .}+\sum_{j}\frac{\left(\sum_{O_{it}=1}\Sigma_{i}\right)\cdot \left(\bv_j \bv_{j}^\top\right)}{\sigma_j^2}\right]\bphi_t - 2\bphi_t^\top\left[(\bV\bigodot \bmu)_{\vec t .}^\top\Lambda^{-1}_{\vec t}\right]\bX_{T,\vec t}\right\}\\
&+\sum_k\lambda_{1k}\bPhi_k^\top\Omega \bPhi_k.
\end{align*}
For a given $k$,  we define the diagonal matrix $\bA\in \real^{T\times T}$ and the vector $a\in \real^T$ as 
\begin{align*}
\bA_{tt}&=\left[(\bV\bigodot \bmu)_{\vec t .}^\top\Lambda^{-1}_{\vec t}(\bV\bigodot \bmu)_{\vec t .}+\sum_{j}\frac{\left(\sum_{O_{it}=1}\Sigma_{i}\right)\cdot \left(\bv_j \bv_{j}^\top\right)}{\sigma_j^2}\right]_{kk}\\
a_t &=(\bV_k\otimes \bmu_{.k})^\top_{\vec t}\Lambda^{-1}_{\vec t}\bX_{T,\vec t}-\left[(\bV\bigodot \bmu)_{\vec t .}^\top\Lambda^{-1}_{\vec t}(\bV\bigodot \bmu)_{\vec t .}+\sum_{j}\frac{\left(\sum_{O_{it}=1}\bSigma_{i}\right)\cdot \left(\bv_j \bv_{j}^\top\right)}{\sigma_j^2}\right]_{k,\bar{k}}^\top\bphi_{t,\bar{k}}.
\end{align*}
When we leave out a specific time point $t_0$, we optimize for $\bPhi_k$ minimizing the following leave-one-out constrained loss,
\begin{align*}
\min_{\bPhi_k}J(t_0, k) &= \bPhi_k^\top \left(\bA(t_0)+\lambda_{1k}\Omega\right)\bPhi_k - 2a(t_0)^\top\bPhi_k, \quad \|\bPhi_k\|_2^2=T.
\end{align*}
We set $\bA(t_0)$ as $\bA$ with the  $(t_0, t_0)$-entry zeroed out, and $a(t_0)$ as $a$ with $a(t_0)$ zeroed out. The leave-one-time cross-validation error is calculated based on the expected deviance loss (unpenalized) at the leave-out time point $t_0$:
\begin{align*}
J_{loocv}(t_0, k) &=\bA_{t_0,t_0}\left(\hat\phi^{-t_0}_{t_0k}  - \frac{a_{t_0}}{\bA_{t_0,t_0}}\right)^2,
\end{align*}
where $\hat\phi^{-t_0}_{t_0k}$ is the leave-one-out estimate. To reduce the computational cost, we drop the norm constraint when picking $\lambda_{1k}$, which enables us to adopt the following short-cut:
\begin{equation}
\label{eq:loocv_shortcut}
\sum_{t_0} J_{loocv}(t_0, k) = \sum_{t_0}\frac{\bA_{t_0,t_0}\left(\hat\phi_{t_0k}  - \frac{a_{t_0}}{\bA_{t_0,t_0}}\right)^2}{\left(1-[(\bA+\lambda_{1k}\Omega)^{-1}]_{t_0 t_0}\right)^2},
\end{equation}
where $\hat\bPhi_k$ is the unconstrained solution from using all time points. This follows from the arguments below:
\begin{enumerate}
\item Drop the constraint, the original problem is equivalent to:
\[
\min \|(\bA(t_0)+\lambda_{1k}\Omega)^{\frac{1}{2}}\bPhi_k-(\bA(t_0)+\lambda_{1k}\Omega)^{-\frac{1}{2}}a(t_0)\|_2^2
\]
\item Consider the augmented problem:
\[
\min \|(\bA+\lambda_{1k}\Omega)^{\frac{1}{2}}\bPhi_k-(\bA+\lambda_{1k}\Omega)^{-\frac{1}{2}}\tilde a(t_0)\|_2^2
\]
where $\tilde a(t_0) = a(t_0)+\delta(t_0)$, and $\delta_{t_0}(t_0) = A_{t0, t_0}\hat\phi_{t_0 k}^{-t_0}$ and $\delta_t(t_0) = 0$ for $t\neq t_0$. The augmented problem also has $\hat\bPhi^{-t_0}_k$ as its solution.
\item Hence, without the constraint, we have $\hat\bPhi_{k} = (\bA+\lambda_{1k}\Omega)^{-1}a$ and $\hat\bPhi_{k}^{-t_0} = (\bA+\lambda_{1k}\Omega)^{-1}\tilde a(t_0)$. Consequently:
\[
\hat\phi_{t_0k} - \hat\phi^{-t_0}_{t_0 k}=[(\bA+\lambda_{1k}\Omega)^{-1}]_{t_0 t_0}(a_{t_0} - \tilde a_{t_0}(t_0))=[(\bA+\lambda_{1k}\Omega)^{-1}]_{t_0 t_0}(a_{t_0} - \bA_{t_0 t_0}\hat\phi_{t_0k}^{-t_0}).
\]
\item Finally, we obtain
\begin{align*}
&\hat \phi_{t_0k} - \frac{a_{t_0}}{\bA_{t_0 t_0}}=\hat \phi_{t_0k}^{-t_0} - \frac{a_{t_0}}{A_{t_0 t_0}}+[(\bA+\lambda_{1k}\Omega)^{-1}]_{t_0 t_0}(a_{t_0} - \bA_{t_0 t_0}\hat\phi_{t_0k}^{-t_0})\\
\Rightarrow&\hat \phi_{t_0k} - \frac{a_{t_0}}{\bA_{t_0 t_0}}=(1-\frac{\bA_{t_0 t_0}}{[(\bA+\lambda_{1k}\Omega)^{-1}]_{t_0 t_0}})(\hat \phi_{t_0k}^{-t_0} - \frac{a_{t_0}}{\bA_{t_0 t_0}})
\end{align*}
Hence, we have$\frac{\bA_{t_0,t_0}\left(\hat\phi_{t_0k}  - \frac{a_{t_0}}{\bA_{t_0,t_0}}\right)^2}{\left(1-[(\bA+\lambda_{1k}\Omega)^{-1}]_{t_0 t_0}\right)^2}=\bA_{t_0,t_0}\left(\hat\phi^{-t_0}_{t_0k}  - \frac{a_{t_0}}{\bA_{t_0,t_0}}\right)^2.$
\end{enumerate}

\subsection{Generation of randomized covariates  for testing}
\label{app:testing_Zgenerate}
In the simulations and real data examples, we encounter two types of $Z_j$: Gaussian and binary. We model the conditional distribution of $Z_j$ given $Z_{j^c}$ with a (penalized) GLM. For Gaussian data, we consider a model where
\[
Z_j = Z_{j^c}\theta+\epsilon_j,\quad \epsilon_j \sim \mathcal{N}(0, \sigma^2).
\]
We estimate $\theta$  and $\sigma^2$ empirically from data. When $q$, the dimension of $Z$, is large, we apply a lasso penalty on $\beta$ with penalty level selected with cross-validation.  Let $\hat \theta$ and $\hat \sigma^2$ be our estimates of the distribution parameters. We then generate new $\bz_i^*$ for subject $i$ from the estimated distribution $\bz_i^* = \bz_{i,j^c}^\top\hat \theta+\epsilon_{ij}^*$, with $\epsilon_{ij}^*$ independently generated from $\mathcal{N}(0,\hat \sigma^2)$. For binary $Z_j$, we consider the model
\[
\log \frac{P(Z_j = 1)}{1-P(Z_j=1)}= Z_{j^c}\theta.
\]
Again, we  estimate $\theta$ empirically, with cross-validated lasso penalty for large $q$. We then generate $z_{ij}^*$ independently from 
\[
P(Z_j = 1|\bz_{j^c}) = \frac{1}{1+\exp(-\bz_{i,j^c}^\top\hat\theta)}.
\]
To generate $Z_j^*$ from the marginal distribution of $Z_j$, instead of estimating this distribution, we let $\bZ_j^*$ be a random permutation of $\bZ_j$.

The randomization test requires generating $Z_j^*$ from the conditional
or marginal distribution of $Z_j$, and estimating the resulting
distribution of $T^*$. We will estimate the
distribution of $T^*$ by fitting a skewed t-distribution as suggested in
\citep{katsevich2020conditional}. The use of the fitted $\hat G(.)$ instead of
the naive empirical CDF can greatly reduce the computational cost: We may obtain accurate estimates of small p-values around $10^{-4}$ using only 200 independent generations of $Z_j^*$ and the fitted $\hat G(.)$.  

\section{Proofs}
\label{app:proofs}
\subsection{Proof of Theorem \ref{thm:monotone}}
\label{app:proof_monotone}
The first step leads to non-decreasing marginal log-likelihood by definition, while the second step is a EM procedure.  If we can show that the penalized marginal log-likelihood is non-decreasing at each subroutine of the EM procedure, we prove Theorem \ref{thm:monotone}. 

For simplicity,  $\theta$ be the parameters that is being updated in some subroutine and $\Theta_{\setminus\theta}$ parameters excluding $\theta$.  Let $\Theta =\Theta_{\setminus\theta}\cup \theta$, $\Theta' =\Theta_{\setminus}\cup \theta$. It is known that if a new $\theta'$ is no worse compared with $\theta$ using the EM objective, it is no worse than $\theta$ when it comes to the (regularized) marginal MLE \citep{dempster1977maximum}.  We include the short proof here for completeness. 

Because the log of the posterior of $\bU$ can be decomposed into the difference between the log of the log  complete likelihood and the log marginal likelihood, $L(\bU|\bX, \Theta) = L(\bU, \bX|\Theta)-L(\bX|\Theta)$, we have (expectation with respect to posterior distribution of $\bU$ with parameters $\Theta$):
\begin{align*}
\bE_{\Theta}L(\bU, \bX|\Theta') = \bE_{\Theta}L(\bX|\Theta')+\bE_{\Theta}L(\bU|\bX, \Theta')=L(\bX| \Theta')+\bE_{\Theta}L(\bU|\bX, \Theta').
\end{align*}
As a result, when 
\[
\bE_{\Theta}L(X, \bU|\Theta')\geq  \bE_{\Theta}L(X, \bU|\Theta),
\]
the following inequality holds,
\begin{align*}
L(\bX| \Theta')-L(\bX|\Theta)&=\{ \bE_{\Theta}L(\bX|\Theta')+\bE_{\Theta}L(\bU|\bX, \Theta')-L(\bX| \Theta')\}+\{\bE_{\Theta}L(\bU|\bX, \Theta')-\bE_{\Theta}L(\bU|\bX, \Theta)\}\\
&\geq  \bE_{\Theta}\log \frac{P(\bU|\bX, \Theta)}{P(\bU|\bX, \Theta')}\geq 0.
\end{align*}
The last inequality is due to the fact that the mutual information $\bE_{\Theta}\log \frac{P(\bU|\bX, \Theta)}{P(\bU|\bX, \Theta')}$ is nonnegative. Now, we return to our subroutines for updating  parameters $(\bV, \bPhi, s^2, \sigma^2)$:
\begin{itemize}
\item  For our subroutines of updating $s^2$ and $\sigma^2$,  they are defined as the maximizers of $\bE_{\Theta}L(X, \bU|\Theta_{-\theta}, \theta)$.
\item From the proof to Lemma \ref{lem:update} in deriving the explicit form for updating $\bPhi$ and $\bV$ (see section\label{app:updateV_Phi_proof1}), we know that $J(\bV_k)$ is a convex quadratic loss and $J(\bV_k) = \sum_{j=1}^J \left(a_{jk}v_{jk}^2+b_{jk}v_{jk}\right)$ for coefficients with explicit forms. If we can show that
\[
\bV_k \leftarrow \min_{V_k} J(\bV_k)+\nu\|\bV_k\|_2^2\quad \|\bV_k\|_2^2=1,
\]
is an minimizer to the problem  $\{\min_{V_k} J(\bV_k),\;s.t.\|\bV_k\|_2^2=1\}$, then updating $\bV_k$ to this new vector is valid and will not increase our loss. This is true by standard optimization arguments. Let $L(\bV_k, \nu)=\min_{\bV_k}\max_{\nu}\left[J(\bV_k)+\nu(\|\bV_k\|_2^2-1)\right]$. Let $\nu$ be such a value such that $\|\bV_k\|_2^2=1$, and let $\bV_k^*$ be the solution at this $\nu$. Then,
\[
J(\bV_k^*) = \min_{\bV_k} L(\bV_k, \nu)\leq  \min_{\bV_k}\max_{\nu} L(\bV_k, \nu)=\min_{\|\bV_k\|_2^2} J(\bV_k)\leq J(\bV_k^*).
\]
Hence, the proposed strategy at line 5 in Algorithm \ref{alg:basicAlg} solves the original problem. The same arguments hold for updating $\bPhi_k$.
\end{itemize}
Hence, given $\Theta^{\ell}$ the posterior distribution at the beginning of iteration $\ell$, if we update the model parameters at proposed to acquire $\Theta^{\ell+1}$, we always have $J(\Theta^{\ell+1})\leq J(\Theta^{\ell})$.

\subsection{Proof of Lemma \ref{lem:initialization}}
\label{app:proof_initialization}
\noindent\fbox{
\parbox{\textwidth}{
\noindent\underline{Statement I:} Suppose  $\bX_{I} = \bU_{\perp}\bM^\top_U$, $\bX_{T} = \bPhi_{\perp}\bM^\top_T$ and   $\bX_{T} = \bV^\top_{\perp}\bM_V$ for some matrices $\bM_U$, $\bM_T$, $\bM_V$. Then,  there exists a rank-K core-array  $\bG=\sum_{k=1}^K \bA_k\circledcirc \bB_k \circledcirc \bC_k$, such that $\bU = \bU_{\perp}\bA$, $\bPhi = \bPhi_{\perp}\bB$ and $\bV = \bV_{\perp}\bC$ form a PARAFAC decomposition of $\bX$:
\[
\bX = \sum_{k=1}^K \bU_{k}\circledcirc \bPhi_k\circledcirc \bV_k.
\] 
\underline{Argument I:} Statement I can be checked easily: since $\bU_{\perp}$ spans the row space of $\bX_{I}$, we can find a matrix $\bA$ such that if we replace $\bU^*$ with $\bU_{\perp}\bA$, we still have a PARAFAC decomposition of $\bX$. We can apply the same arguments to $\bPhi_{\perp}, \bV_{\perp}$, and hence prove the statement at the beginning.
}
}
As a result, we need to show that $\bU_{\perp}$, $\bPhi_{\perp}$ and $\bV_{\perp}$ spans the column spaces of the three unfolding matrices and $\tilde \bG$ is the unfolding of such a core array along the subject dimension.
\begin{itemize}
\item The proposed $\bV_{\perp}$ satisfies the requirement by construction. Hence, we need only to check that $\bU_{\perp}$ and $\bPhi_{\perp}$ spans the row space of $\bX_{I}$ and $\bX_T$ respectively.
\item  The projection of $\bX_J$ onto $\bV_{\perp}$ results in $H = \bV_{\perp}^\top\bX_j = \bC (\bPhi^*\odot \bU^*)^\top$, where $\bC = \bV_{\perp}^\top\bV^*$. Hence, we have 
\[
\bW =  \sum_{k=1}^K \frac{1}{I}\bY(k)^\top\bY(k) =\bPhi^* \left(\sum_{k=1}^K\left(\bU^*\cdot \bC_{k,.}\right)^\top\left(\bU^*\cdot \bC_{k,.}\right) \right)\bPhi^{*,T}=\bPhi^* \bM\bPhi^{*\top}
\]
where $\bM = \real^{K\times K}$ with $\bM_{\ell k} =\langle\bC_{\ell},\bC_k \rangle\langle\bU^*_{\ell},\bU^*_k \rangle$.  Notice that
\begin{align*}
\bX_{T}\bX_{T}^\top &= \bPhi^* \left(\bV^*\odot \bU^*\right)^\top\left(\bV^*\odot \bU^*\right)\bPhi^{*,T}\\
& = \bPhi^* \left((\bV_{\perp}\bC_{1})\otimes \bU^*_1,\ldots, (\bV_{\perp}\bC_{K})\otimes \bU^*_K \right)^\top \left((\bV_{\perp}\bC_{1})\otimes \bU^*_1,\ldots, (\bV_{\perp}\bC_{K})\otimes \bU^*_K \right)\bPhi^{*\top}\\
& = \bPhi^* \left(\left(\bC^\top_{\ell}\bV^\top_{\perp}\bV_{\perp}\bC_{k} \right)\otimes(\bU_\ell^\top\bU_k)\right)_{\ell k}\bPhi^{*\top}\\
& =  \bPhi^* \left(\left(\bC^\top_{\ell}\bC_{k} \right)(\bU_\ell^\top\bU_k)\right)_{\ell k}\bPhi^{*\top} = \bPhi^* \bM\bPhi^{*\top}
\end{align*}
Hence, $\Phi_{\perp}$ is the top $K$ left singular vectors of $\bX_{T}$. We set $\bB = \bPhi_{\perp}^\top \bPhi^*$.
\item  $\tilde\bU$ is estimated by regression $\bX_{I}$ on $\bV_{\perp}\otimes \bPhi$. Because both $\bV_{\perp}$ and $\bPhi_{\perp}$ are orthonormal, $\bV_{\perp}\otimes \bPhi_{\perp}$ is also orthonormal:
\[
(\bV_{\perp}\otimes \bPhi)^\top(\bV_{\perp}\otimes \bPhi) = (\bV_{\perp}^\top\bV_{\perp})\otimes (\bPhi^\top\bPhi_{\perp})= \Id_{K\times K}\otimes \Id_{K\times K} = \Id_{K^2\times K^2}.
\]
Hence,  we have
\begin{align*}
\tilde\bU (\bV_{\perp}\otimes \bPhi_{\perp})^\top&=\frac{1}{1+\delta} \bX_{I}(\bV_{\perp}\otimes \bPhi_{\perp}) (\bV_{\perp}\otimes \bPhi_{\perp})^\top\\
&=\frac{1}{1+\delta} \bX_{I}((\bV_{\perp}\bV_{\perp}^\top)\otimes (\bPhi_{\perp}\bPhi_{\perp}^\top))\\
& =\frac{1}{1+\delta} \bU^* \left((\bV_{\perp} \bC) \odot (\bPhi_{\perp} \bB)\right)^\top(\bV_{\perp}\otimes \bPhi_{\perp})(\bV_{\perp}\otimes \bPhi_{\perp})^\top\\
& = \bU^*\left( \begin{array}{c} (\bC^\top_1\bV^\top_{\perp}\bV_{\perp}\bV^\top_{\perp}) \otimes (\bB_1^\top\bPhi^\top_{\perp}\bPhi_{\perp}\bPhi^\top_{\perp})\\\ldots \\(\bC^\top_K\bV^\top_{\perp}\bV^\top_{\perp}\bV^\top_{\perp}) \otimes (\bB_K^\top\bPhi^\top_{\perp}\bPhi_{\perp}\bPhi^\top_{\perp})\\ \end{array}\right)(\bV_{\perp}\otimes \bPhi_{\perp}) \\
& = \frac{1}{1+\delta}\bX_{I}
\end{align*}
The row space spanned by $\tilde\bU$ is the same as the row space spanned by $\bX_{I}$, thus, the  space spanned by top $K$ left singular vectors of $\tilde \bU$ is the same by that of $\bX_{I}$. As a result, $\bU_{\perp}$ also satisfies the requirement. 

\item In particular, we also have
\begin{align*}
\bU_{\perp}^\top\tilde \bU& = \frac{1}{1+\delta}\bU_{\perp}^\top\bX_{I} (\bV_{\perp}\otimes \bPhi_{\perp}) = \frac{1}{1+\delta}\bU_{\perp}^\top \bU_{\perp}\bG_{I}(\bV_{\perp}\otimes \bPhi_{\perp})^\top(\bV_{\perp}\otimes \bPhi_{\perp}) = \bG_{I}
\end{align*}
where $\bG_{I}$ is the unfolding of the core array $\bG$ in the subject dimension. Hence, we recover $\bA, \bB, \bC$ applying a rank-K PARAFAC decomposition on the arranged three-dimensional core array from $\bG_{I}$ as described in Algorithm \ref{alg:initialization}.
\end{itemize}
\subsection{Proof of Lemma \ref{lem:testing1}}
\label{app:proof_testing}
Let  $\boldsymbol{e}_k = (\underbrace{0,\ldots,0}_{k-1}, 1,\underbrace{0,\ldots,0}_{K-k})^\top$. Plug-in the expression of $\bX_{I,\vec i} = (\bV\bigodot \bPhi)_{\vec i}\bu_i=(\bV\bigodot \bPhi)_{\vec i}((\bbeta^*)^\top\bz_i+\zeta_i)+\epsilon_{I,\vec i}$ into the expression of $\tilde y_i(\delta)$:
\begin{align*}
\tilde y_i(\delta) &= \left((\bSigma_{i}(\delta))_{k:}(\bV\odot \bPhi)^\top\Lambda_{\vec i}\bX_{I,\vec i}+\delta\sum_{\ell\neq k}\frac{(\bSigma_{i}(\delta))_{kl}}{s_{\ell}^2}\bz_i^\top\bbeta_{\ell}\right)\\
& =  (\boldsymbol{e}_k^\top  - \delta(\bSigma_{i}(\delta))_{k:}\Lambda_f)(\bbeta^*)^{\top}\bz_i+\delta\sum_{\ell\neq k}\frac{(\bSigma_{i}(\delta))_{kl}}{s_{\ell}^2}\bz_i^\top\bbeta_{\ell}+(\boldsymbol{e}_k^\top -\delta(\bSigma_{i}(\delta))_{k:}\Lambda_f)\zeta_i\\
&+(\bSigma_{i}(\delta))_{k:} (\bV\bigodot\bPhi)^\top_{\vec i}\Lambda_{\vec i}\epsilon_{I,\vec i}\\
& = (1-\delta\frac{(\bSigma_{i}(\delta))_{kk}}{s_k^2})\bz_i^\top\bbeta_k^*+\underbrace{\sum_{\ell\neq k}\frac{(\bSigma_{i}(\delta))_{kl}}{s_l^2}\bz_i^\top(\bbeta_{\ell}-\bbeta_{\ell}^*)}_{\Delta_i(\delta)}+\xi_i
\end{align*}
where $\xi_i = (\boldsymbol{e}_k^\top -\delta(\bSigma_{i}(\delta))_{k:}\Lambda_f)\zeta_i+(\bSigma_{i}(\delta))_{k:}(\bV\bigodot\bPhi)^\top_{\vec i}\Lambda_{\vec i}\epsilon_{I,\vec i}$ and
\begin{align*}
\bE(\xi_i^2)& = (\boldsymbol{e}^\top_k -\delta(\bSigma_{i}(\delta))_{k:} \Lambda_f)\Lambda_f^{-1} (\boldsymbol{e}_k -\delta\Lambda_f(\bSigma_{i}(\delta))_{:k})+(\bSigma_{i}(\delta))_{k:}(\bV\bigodot\bPhi)^\top_{\vec i}\Lambda_{\vec i}(\bV\bigodot\bPhi)^\top_{\vec i}(\bSigma_{i}(\delta))_{:k}\\
& = s_k^2 -2\delta(\bSigma_{i}(\delta))_{kk} +\delta^2(\bSigma_{i}(\delta))_{k:}\Lambda_f(\bSigma_{i}(\delta))_{:k}(\delta)+(\bSigma_{i}(\delta))_{k:}(\bSigma_i^{-1}-\delta\Lambda_f)(\bSigma_{i}(\delta))_{:k}\\
& = s_k^2 +(1-2\delta)(\bSigma_{i}(\delta))_{kk}(\delta)+(\delta^2-\delta)(\bSigma_{i}(\delta))_{k:}\Lambda_f(\bSigma_{i}(\delta))_{:k} = w_i(\delta)
\end{align*}

\subsection{Proof of Lemma \ref{lem:update}}
\label{app:proof_update}
\subsubsection{Update of $\bbeta_k$:} 
From eq.\ (\ref{eq:loglik_marginal2}) and eq.\ (\ref{eq:objective}), update of $\bbeta$ considers the objective:
\begin{align}
\label{eq:update_beta}
&J(\bbeta)\\
=&\frac{1}{2}\boldf_i^\top\left( \Lambda_{\vec i}^{-1}- \Lambda_{\vec i}^{-1}\bH_{\vec i}\Sigma_i \bH_{\vec i}^\top \Lambda_{\vec i}^{-1}\right)\boldf_i+\sum_k\lambda_{2k}|\bbeta_k|\notag\\
=&\frac{1}{2}\left[\deta_i^\top M^i \deta_i+ \bX_{I\vec i}^\top  \left( \Lambda_{\vec i}^{-1}- \Lambda_{\vec i}^{-1}\bH_{\vec i}\Sigma_i \bH_{\vec i}^\top \Lambda_{\vec i}^{-1}\right) \bX_{I\vec i}+2\deta_i^\top m^i\right]+\sum_k\lambda_{2k}|\bbeta_k|\notag,
\end{align}
where
\begin{align}
\deta_i&=\bbeta^\top\bz_i ,\notag\\
M^i & = \Lambda_f^{-1}-\Lambda_f^{-1}\Sigma_i \Lambda_f^{-1},\notag\\
m^i & = \Lambda_f^{-1} \Sigma_i \bH_{\vec i}^\top \Lambda_{\vec i}^{-1}\bX_{I\vec i}.\notag
\end{align}
Hence, for $\bbeta_k$, we consider the following minimization problem
\begin{align*}
J(\beta_k)&=\frac{1}{2}\left[\beta_k^\top z_i^\top M^i_{kk} z_i\beta_k-2\beta_k^\top z_i^\top \left(m^i_k-\sum_{\ell\neq k}M^i_{k\ell} z_i \beta_{\ell}\right)\right]+\lambda_{2k}|\bbeta_k|,\\
& = \beta_k^\top \tilde z_i^\top  \tilde z_i\beta_k-2\beta_k^\top \tilde z_i^\top \underbrace{\left(M_{kk}^i\right)^{-\frac{1}{2}}\left(m^i_k-\sum_{\ell\neq k}M^i_{k\ell} z_i \beta_{\ell}\right)}_{\tilde y_i}.
\end{align*}
It is straightforward to check that the definitions of $\tilde z_i$ and $\tilde y_i$ are the same as the ones used in Lemma \ref{lem:update}.

\subsubsection{Update of $\bV_k$ and $\bPhi_k$:}
\label{app:updateV_Phi_proof1}
We can rewrite the complete log-likelihood using the unfolding along the feature dimension and the time dimension. The relevant terms to $\bV$ and $\bPhi$ are:
\begin{align*}
L(\bX,\bU|\Theta)=&-\frac{1}{2}\sum_j \left(\bX_{J,\vec j}-(\bPhi\odot\bU)_{\vec
j}\bv_j \right)^\top\Lambda_{\vec j}^{-1} \left(\bX_{J,\vec j}-(\bPhi\odot\bU)_{\vec
j}\bv_j \right),\\
=&-\frac{1}{2}\sum_t \left(\bX_{T,\vec t}-(\bV\odot\bU)_{\vec
j}\bphi_j \right)^\top\Lambda_{\vec t}^{-1} \left(\bX_{J,\vec j}-(\bV\odot\bU)_{\vec
j}\bphi_j \right).
\end{align*}
Let $O_{it}=1$ if observation time index $t$ for subject $i$ is available, and let $o_i$ be the set of index where $O_{it}=1$ for subject i.  By direct calculation, the expectation of the negative log-likelihood (keeping only terms relevant to $\bPhi$ and $\bV$ ) are given below:
\begin{align}
\label{eq:expected_V}
&\frac{1}{2}\bE[\sum_j \left(\bX_{J,\vec j}-(\bPhi\odot\bU)_{\vec
j}\bv_j \right)^\top\Lambda_{\vec j}^{-1} \left(\bX_{J,\vec j}-(\bPhi\odot\bU)_{\vec
j}\bv_j \right)],\\
\propto&\frac{1}{2}\sum_j \sum_i\left(-2\bX_{ij,o_{i}}^\top\Lambda_{\vec j}^{-1} \left(\bPhi_{o_i.}\odot\langle\bu_{i.}\rangle\right)\bv_j+\bv_j^\top\ \left(\bPhi_{o_i.}^\top \Lambda_{\vec j}^{-1}\bPhi_{o_i.}\cdot\langle\bu_i\bu_i^\top \rangle \right)\bv_{j}\right).\notag
\end{align}

\begin{align}
\label{eq:expected_Phi}
&\frac{1}{2}\bE[\sum_t \left(\bX_{T,\vec t}-(\bV\odot\bU)_{\vec
j}\bphi_t \right)^\top\Lambda_{\vec t}^{-1} \left(\bX_{T,\vec t}-(\bV\odot\bU)_{\vec
t}\bphi_t  \right)],\\
\propto&\frac{1}{2}\sum_t \sum_{i:O_{it}=1}\left(-2\bX_{i, .,t}^\top\Lambda_{\vec t}^{-1} \left(\bV\odot\langle\bu_{i.}\rangle\right)\bphi_j+\bphi_t^\top\ \left(\bV^\top \Lambda_{\vec t}^{-1}\bV\cdot\langle\bu_i\bu_i^\top \rangle \right)\bphi_{t}\right).\notag
\end{align}
\noindent\textbf{Update $V_k$: }
From eq.\ (\ref{eq:expected_V}), update of $\bV_k$ considers the following problem
\begin{align*}
\min_{\bV_k}J(\bV_k)&= \sum_{j=1}^J\frac{1}{2} \left(a_{jk} v_{jk}^2 -2b_{jk} v_{jk}\right), \quad s.t. \quad \|\bV_k\|_2^2 = 1.
\end{align*}
Here,  $a_{jk}$ and $b_{jk}$ are defined below:
\begin{align*}
a_{jk} & =\frac{1}{\sigma_j^2} \sum_{i,t:O_{it}=1}\left(\phi_{tk}^2 \langle u_{ik}\rangle^2+\Sigma^i_{kk}\phi_{tk}^2\right),\\
& =\sum_{i,t:O_{it}=1}\left(\phi_{tk}^2 \langle u^2_{ik}\rangle\right) = \langle \|(\bU\odot \Phi)_{\vec j, k}\|_2^2\rangle.\\
b_{jk}& = \frac{1}{\sigma_j^2}  \sum_{i,t:O_{itj}=1}\left(\left(x_{ijt} - \sum_{k'\neq k} v_{jk'}\phi_{tk'}\langle \bu_{ik'}\rangle\right)\langle \bu_{ik}\rangle\phi_{tk}-\sum_{k'\neq k}\Sigma_{kk'}^i v_{jk'}\phi_{tk'}\phi_{tk}\right).
\end{align*}
Hence, if we set $Q_1$, $Q_2$, $Q_3$ as specified, we have $a_{jk} = \frac{1}{\sigma_j^2}\|Q_{1,\vec j, k}\|_2^2+Q_{3kk,j}$ and $b_{jk} = Q_{2kj} - \sum_{k'\neq k}Q_{3kk'}v_{jk'}$.

\noindent\textbf{Update $\bPhi_k$}
Updating $\bPhi_k$ is similar to updating $\bV_k$ apart from the inclusion of the smoothness regularizer. Set  $Q_1=\langle\bU\rangle\odot\bPhi\in \real^{IT\times K}$,  $Q_2\in \real^{K\times T}$ and $Q_3\in \real^{K\times K\times T}$ with
\begin{align*}
Q_{2kt}& = \bX_{Tt\vec t}^\top\diag\{\frac{1}{\sigma_{\vec t}^2}\} Q_{1\vec t k}\\
Q_{3k_1k_2 t}&=\sum_{j}\frac{1}{\sigma_j^2}v_{jk_1}v_{jk_2}\sum_{i:O_{it}=1}\left(\Sigma_{ik_1 k_2}+\langle u_{ik_1}\rangle\langle u_{ik_2}\rangle \right)\end{align*}
Then,
\begin{align*}
J(\bPhi_k)&=\sum_{t=1}^T\frac{1}{2} \left(a_{jk} v_{jk}^2 -2b_{jk} v_{jk}\right)+\lambda_{1k}\bPhi_k^\top\Omega \bPhi_k,
\end{align*}
where $a_{tk} =\left( \|\diag\{\frac{1}{\sigma_{\vec t}}\}Q_{1,\vec t, k}\|_2^2+Q_{3kk,t}\right)$ and $b_{jk} =\left(Q_{2kt} - \sum_{k'\neq k}Q_{3kk'}\phi_{tk'}\right)$.
\subsubsection{Update of $\sigma_j^2$ and $s_k^2$}
Since the expected log likelihood related to $\{s^2_k,\;k=1,\ldots,K\}$ is
\begin{align*}
& \frac{1}{2}\sum_{i=1}^I\{\bE_{\Theta_0}[-(\bu_i - \bbeta^\top\bz_i)^\top\Lambda_f^{-1} (\bu_i - \bbeta^\top\bz_i)]-\log |\Lambda_f|\}\\
=&  \sum_{k=1}^K-\left\{\frac{\sum_{i=1}^I\left((\mu_{ik}-\bz_i^\top\bbeta_k)^2+(\bSigma_i)_{kk}\right)}{s_k^2}+I\log s_k^2\right\}
\end{align*}
Consequently, the solution for $s^2_k$ given other parameters is  $s^2_k= \frac{1}{I}\sum^{I}_{i=1}\left((\mu_{ik}-\bz_i^\top\bbeta_k)^2+(\bSigma_i)_{kk}\right)$.

the expected log likelihood related to $\{\sigma^2_j,\;j=1,\ldots,p\}$ is
\begin{align*}
-\frac{1}{2}\sum^J_{j=1}\left\{\frac{1}{\sigma_j^2}\left[(X_{J,\vec j}-(\bPhi\bigodot \mu)_{\vec j} \bv_{j})^\top(X_{J,\vec j}-(\bPhi\bigodot \mu)_{\vec j} \bv_{j})+\bv_j^\top\left(\sum_{t}(\sum_{O_{it}=1}\bSigma_{i})\cdot (\bphi_{t}\bphi_t^\top)\right)\bv_j\right]+|\vec j|\log \sigma_j^2\right\}
\end{align*}
Consequently, the updating rule $\sigma^2_h$ given other parameters is  
\[
 \sigma_j^2= \frac{1}{|\vec j|}\left[(X_{J,\vec j}-(\bPhi\bigodot \mu)_{\vec j} \bv_{j})^\top(X_{J,\vec j}-(\bPhi\bigodot \mu)_{\vec j} \bv_{j})+\bv_j^\top\left(\sum_{t}(\sum_{O_{it}=1}\bSigma_{i})\cdot (\bphi_{t}\bphi_t^\top)\right)\bv_j\right].
\] 
\section{Additional numerical results}
\subsection{Comparison of two initialization strategies}
\label{app:numerical_init}
\begin{figure}[H]
\includegraphics[width = 1\textwidth, height =0.6\textwidth]{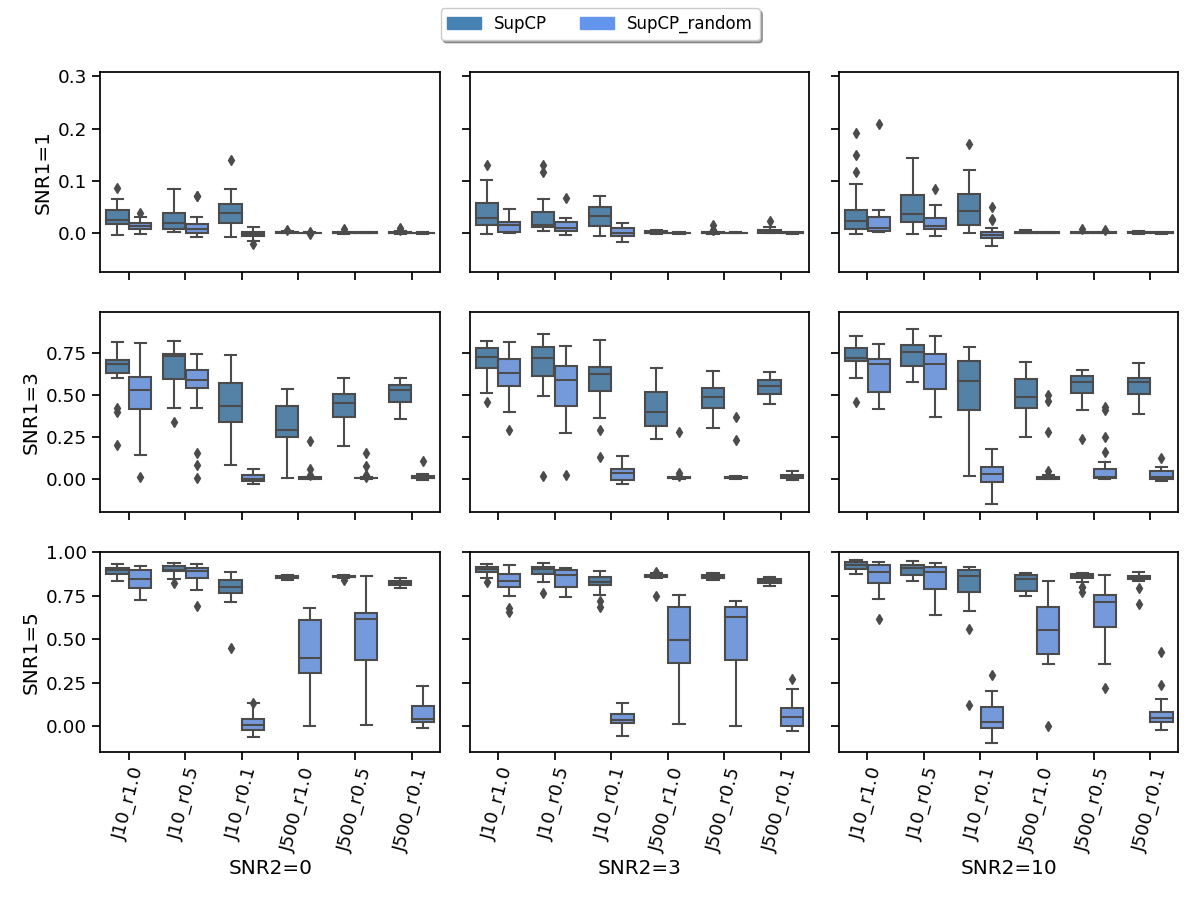}
\caption{Reconstruction evaluation of the underlying signal tensor using different initializations in SupCP, measured by the correlations. In each subplot, x axis label indicates different $J$ and observing rate,  the y axis is the achieved correlation, and  the box colors represent different methods. The  corresponding subplot column/row name represents the signal-to-noise ratio SNR1\slash SNR2.}
\label{fig:construction_init}
\end{figure}

Fig.\ref{fig:construction_init} compares the achieved correlations with the signal tensor when SupCP are initialized using the proposed initialization method (referred to as SupCP) and the random initialization method (referred to as SupCP\_random)  \citep{lock2018supervised}. The proposed strategy shows a clear gain in the setting of high missing rate or weak signal.
\subsection{Signal reconstruction on missing entries}
\label{app:numerical_reconstruct}
\begin{figure}[H]
\includegraphics[width  = 1.0\textwidth]{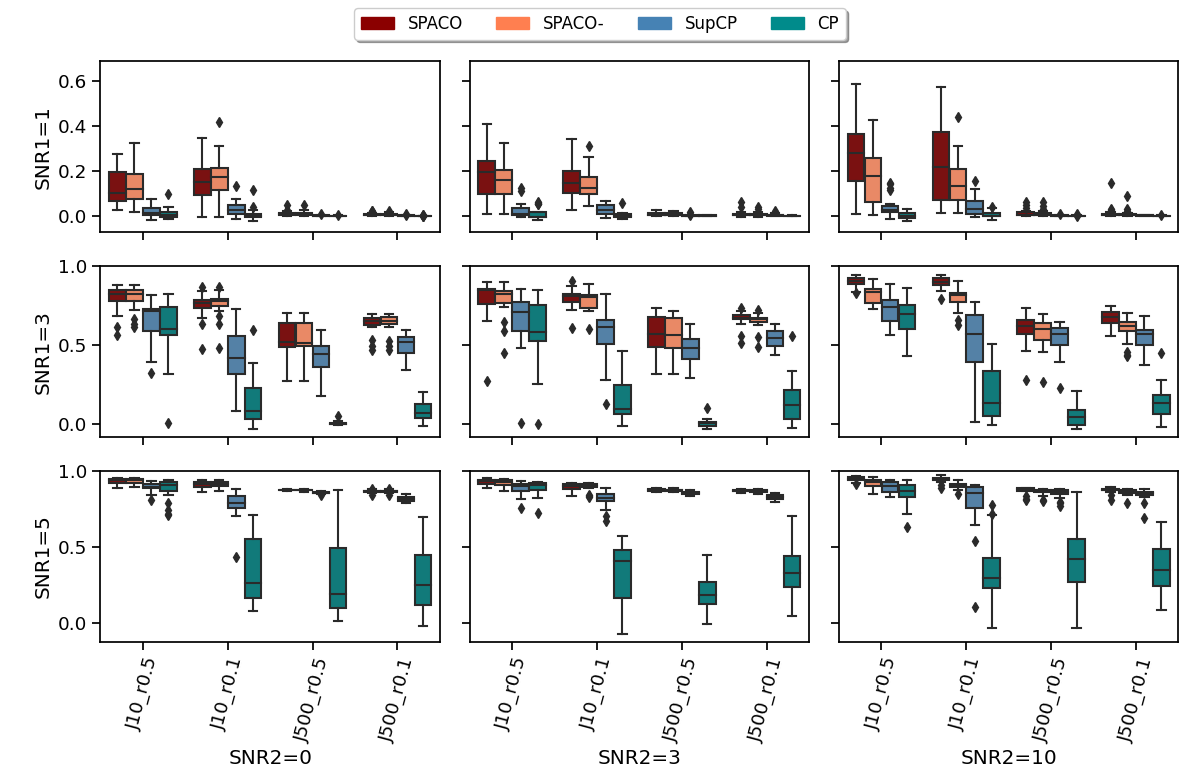}
\caption{Reconstruction evaluation on missing entries by the correlations between the estimates and the true signal tensor. In each subplot, x axis label indicates different $J$ and observing rate,  the y axis is the achieved correlation, and  the box colors represent different methods. The  corresponding subplot column/row name represents the signal-to-noise ratio SNR1\slash SNR2.}
\label{fig:construction_signal_miss}
\end{figure}

Fig.\ref{fig:construction_signal_miss} compares the achieved correlations with the signal tensor using different methods, but only on those missing entries. The proposed strategy shows a clear gain in the setting of high missing rate or weak signal. Consistent with Fig.\ref{fig:construction_signalmat}, SPACO improves over SPACO-  with high SNR2, and is much better than SupCP or CP when the signal size SNR1 is weak (but still estimable using SPACO) or when the missing rate is high.

\subsection{Type I error with observing rate $r = 1.0, 0.1$}
\label{app:numerical_typeI}
\begin{figure}[H]
\begin{center}
\caption{Achieved type I errors at observing rate $r = 1.0$. In each subplot, x axis label indicates different combination of feature dimension $J$ and targeted level $\alpha\in \{0.01, 0.05\}$,  the y axis is the achieved type I errors. Different bar colors represent different tests (partial or marginal). The two dashed horizontal lines indicate levels 0.01 and 0.05.}
\label{fig:typeI_main}
\includegraphics[width  = .9\textwidth, height = .67\textwidth]{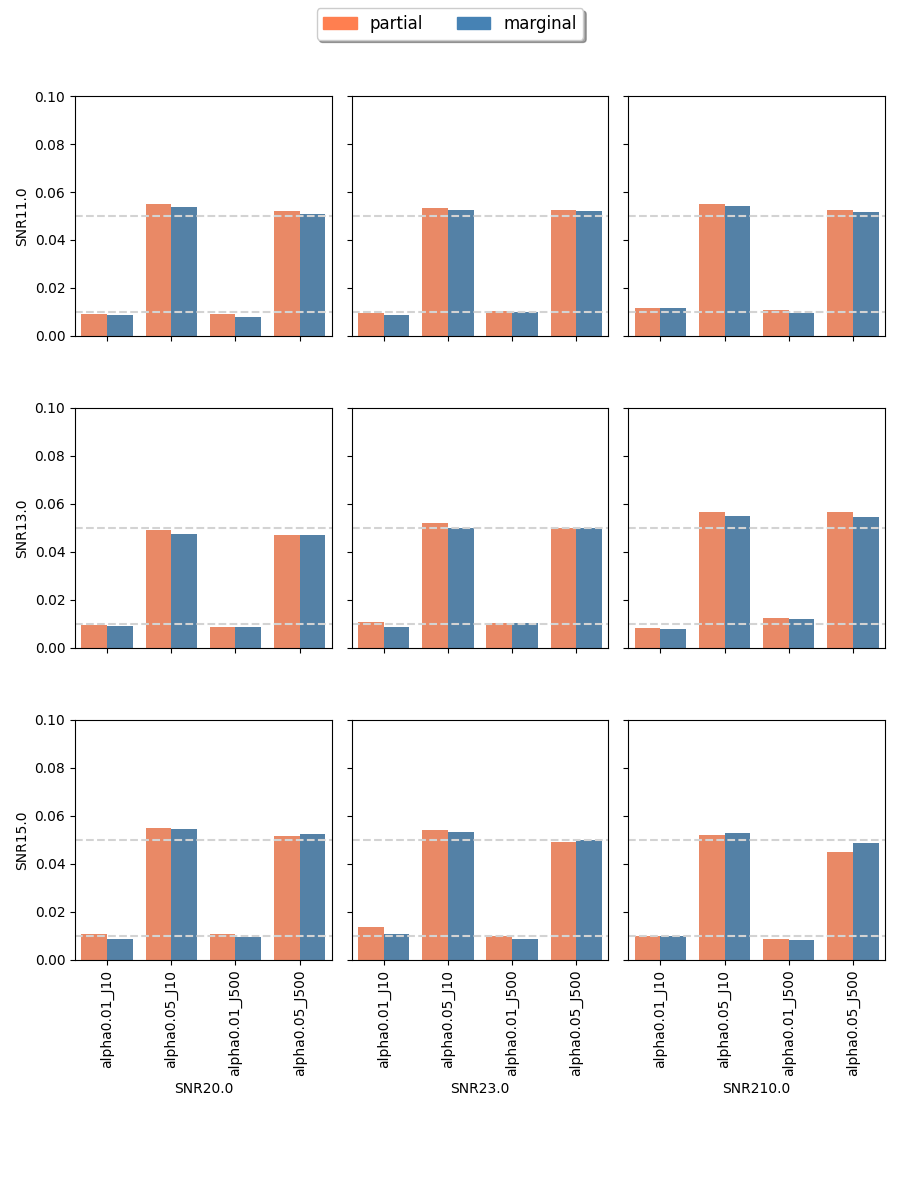}
\end{center}
\end{figure}

\begin{figure}[H]
\begin{center}
\caption{Achieved type I errors at observing rate $r = 0.1$. In each subplot, x axis label indicates different combination of feature dimension $J$ and targeted level $\alpha\in \{0.01, 0.05\}$,  the y axis is the achieved type I errors. Different bar colors represent different tests (partial or marginal). The two dashed horizontal lines indicate levels 0.01 and 0.05.}
\label{fig:typeI_main}
\includegraphics[width  = .9\textwidth, height = .67\textwidth]{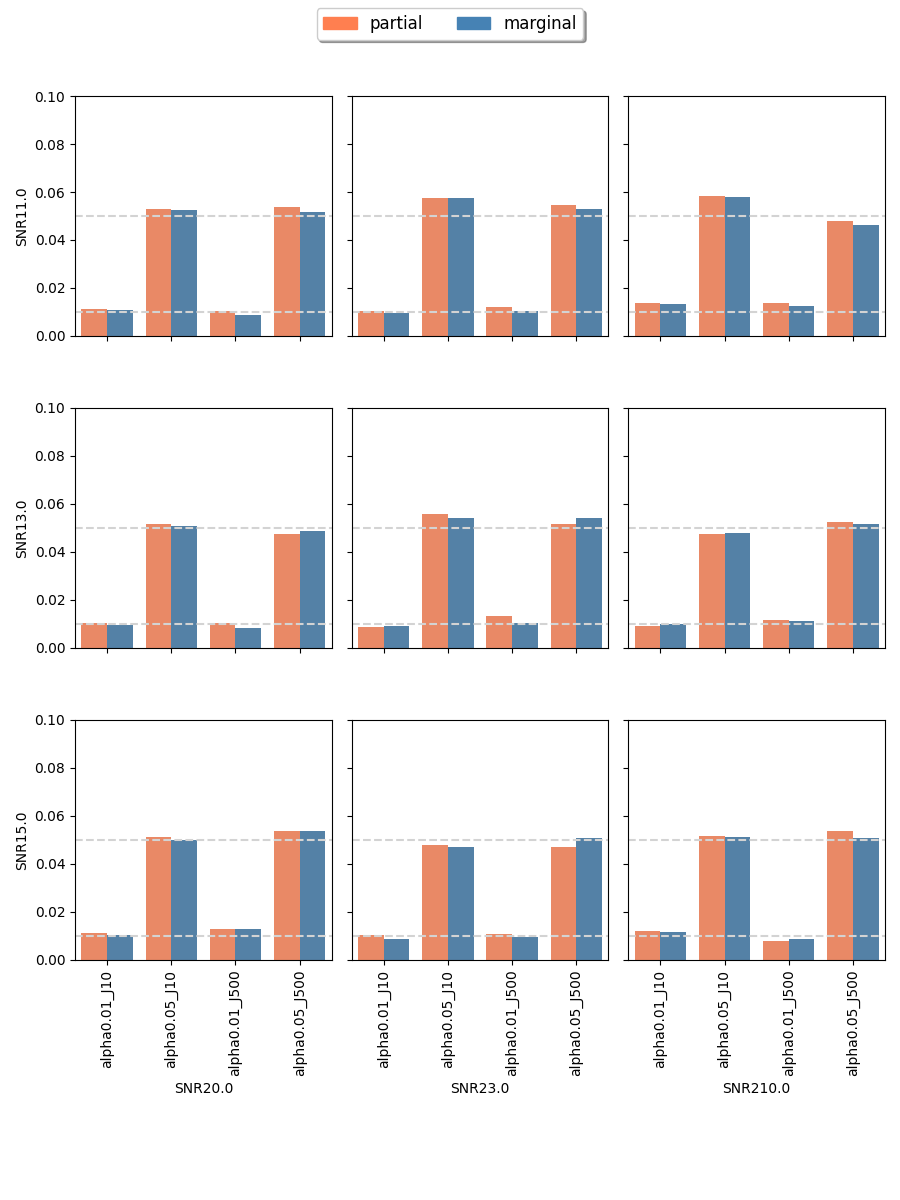}
\end{center}
\end{figure}

\subsection{empirical p-value: cross-fit vs naive fit}
\label{app:cross_naive_pval}
\begin{figure}[H]
\caption{qq-plots of constructed p-values at $r = 1.0$. The p values are estimated with nct distribution. The red/blue points represent those from the cross-fit and the naive procedure, with blue diagonal representing the expected theoretical behavior.}
\label{fig:qq_crossfitI}
\begin{center}
\includegraphics[width  =0.49\textwidth]{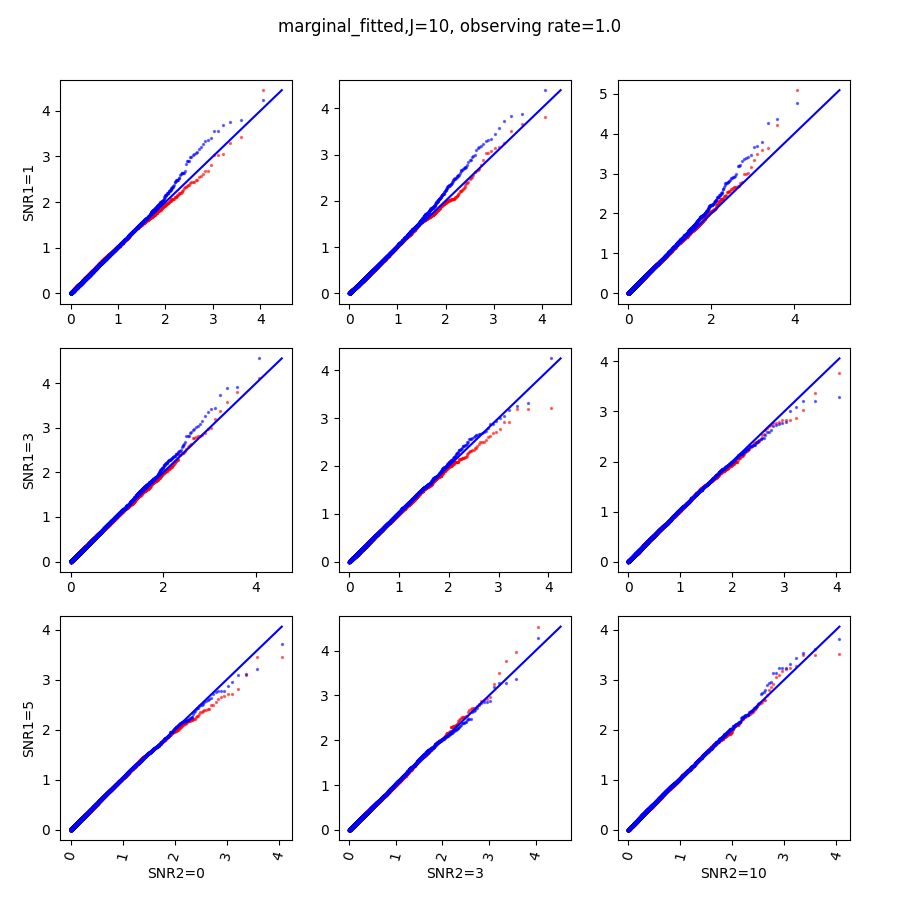}
\includegraphics[width  =0.49\textwidth]{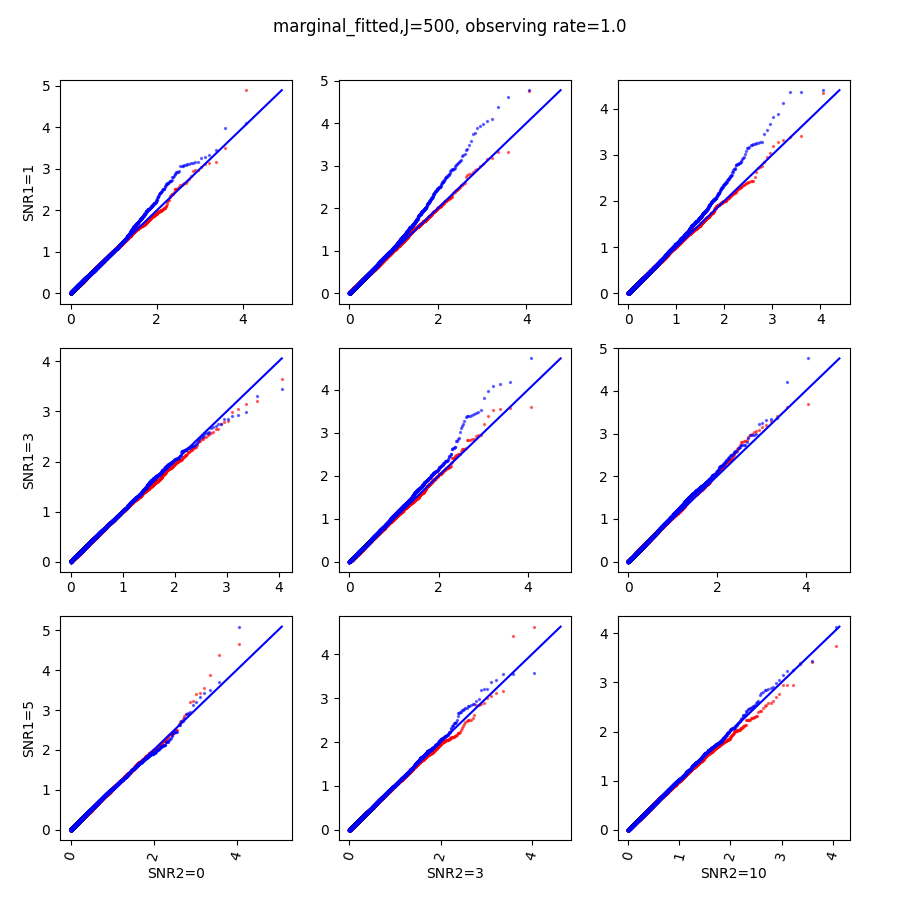}\\
\includegraphics[width  =0.49\textwidth]{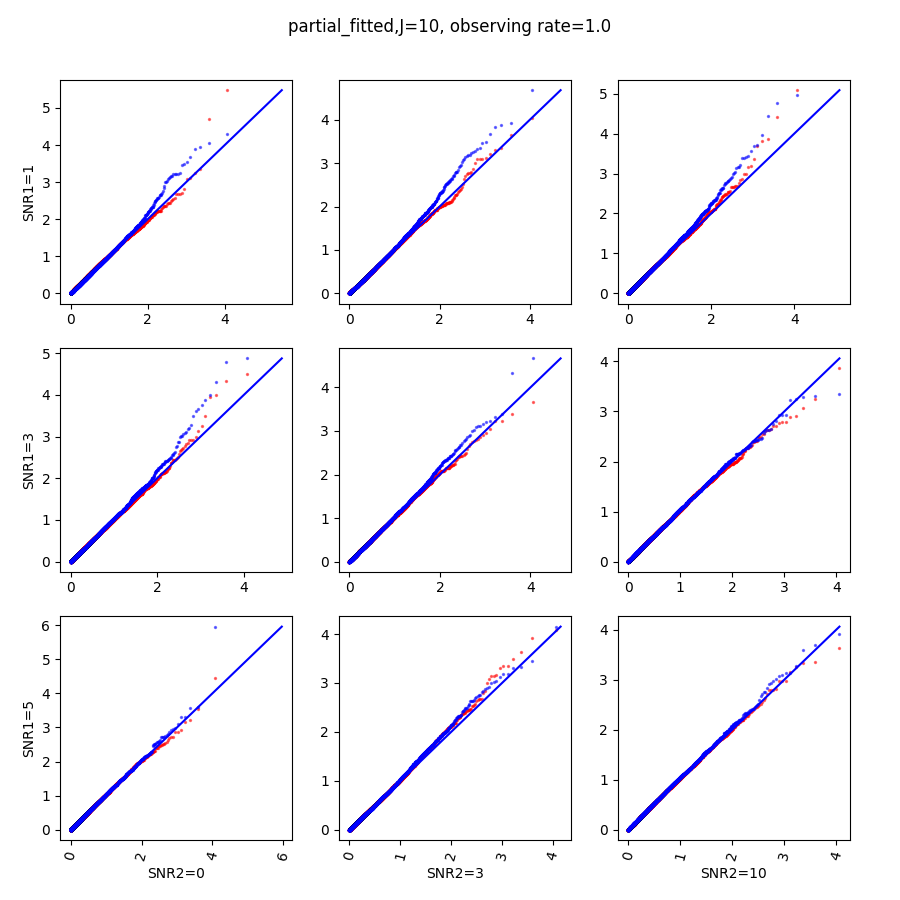}
\includegraphics[width  =0.49\textwidth]{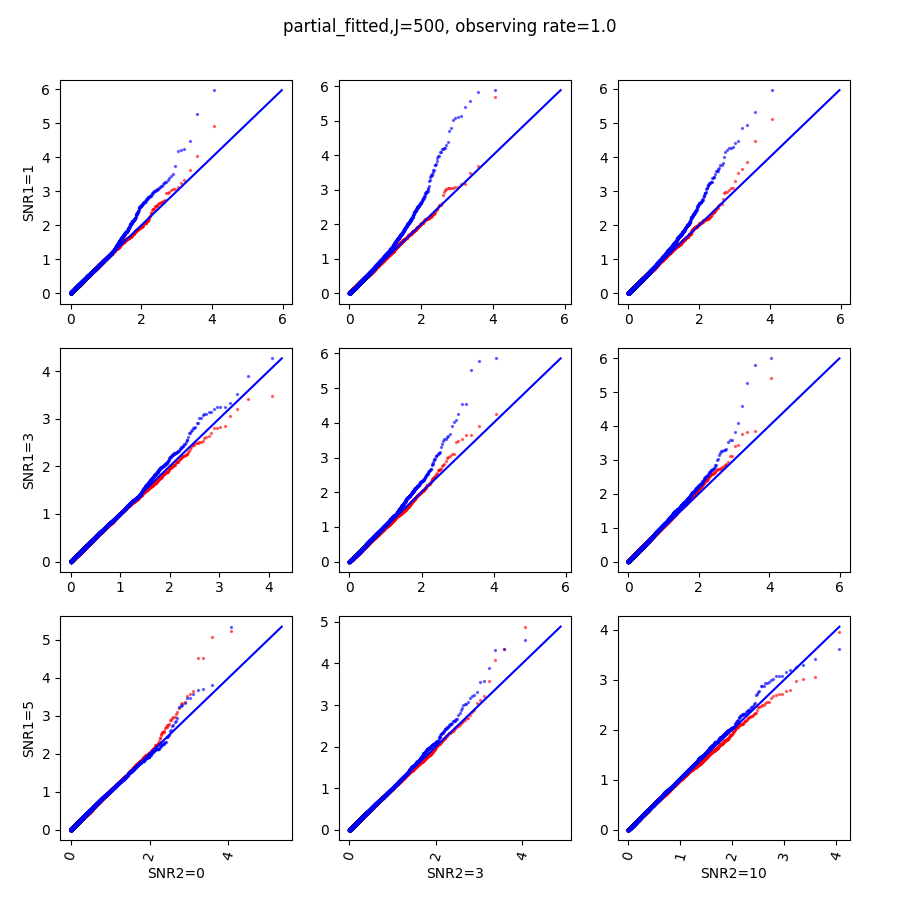}
\end{center}
\end{figure}

\begin{figure}[H]
\caption{qq-plots of constructed p-values at $r = 0.5$. The p values are estimated with nct distribution. The red/blue points represent those from the cross-fit and the naive procedure, with blue diagonal representing the expected theoretical behavior.}
\label{fig:qq_crossfitII}
\begin{center}
\includegraphics[width  =0.49\textwidth]{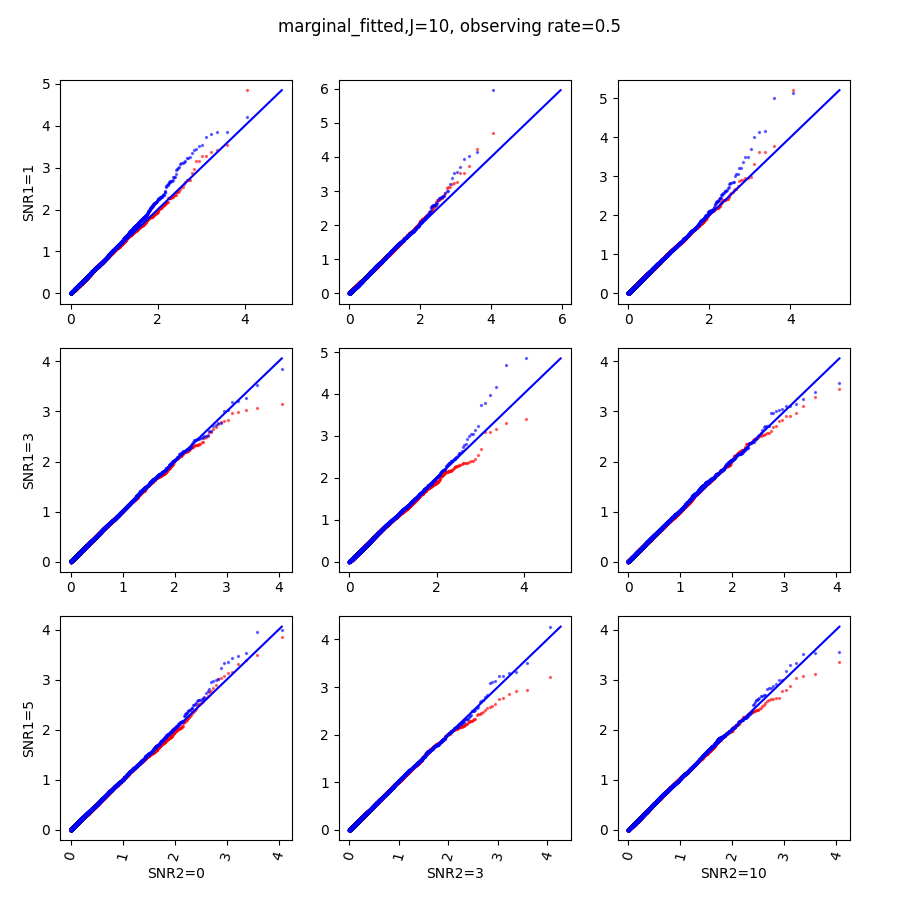}
\includegraphics[width  =0.49\textwidth]{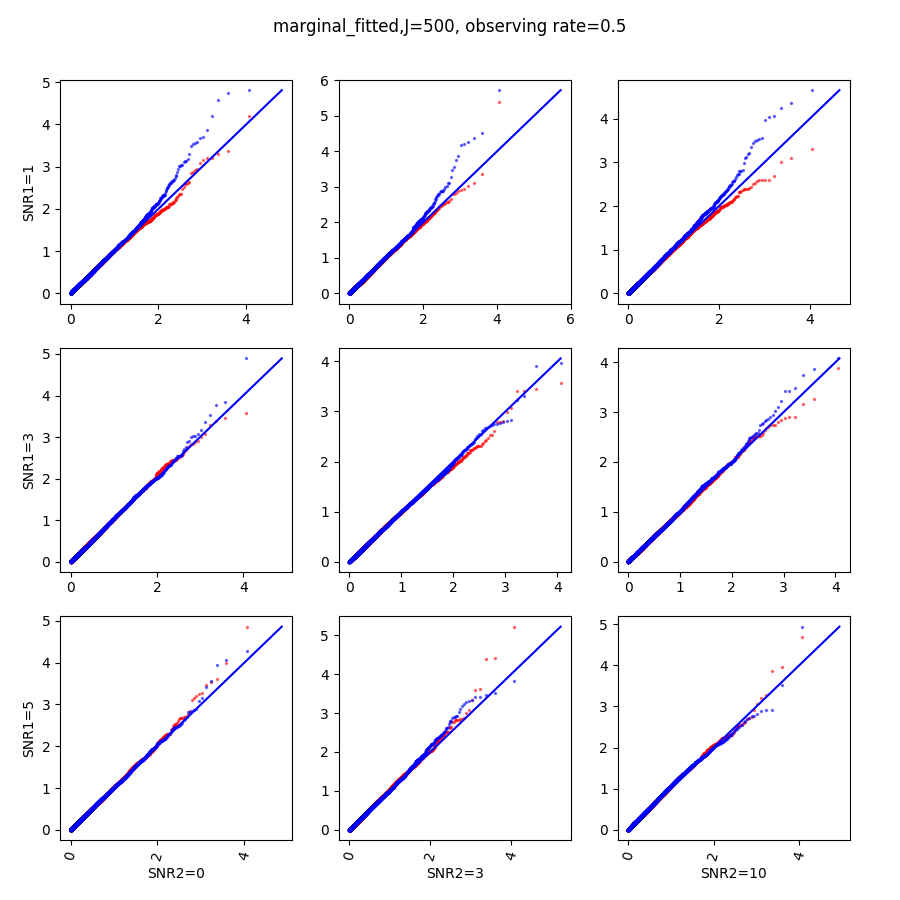}\\
\includegraphics[width  =0.49\textwidth]{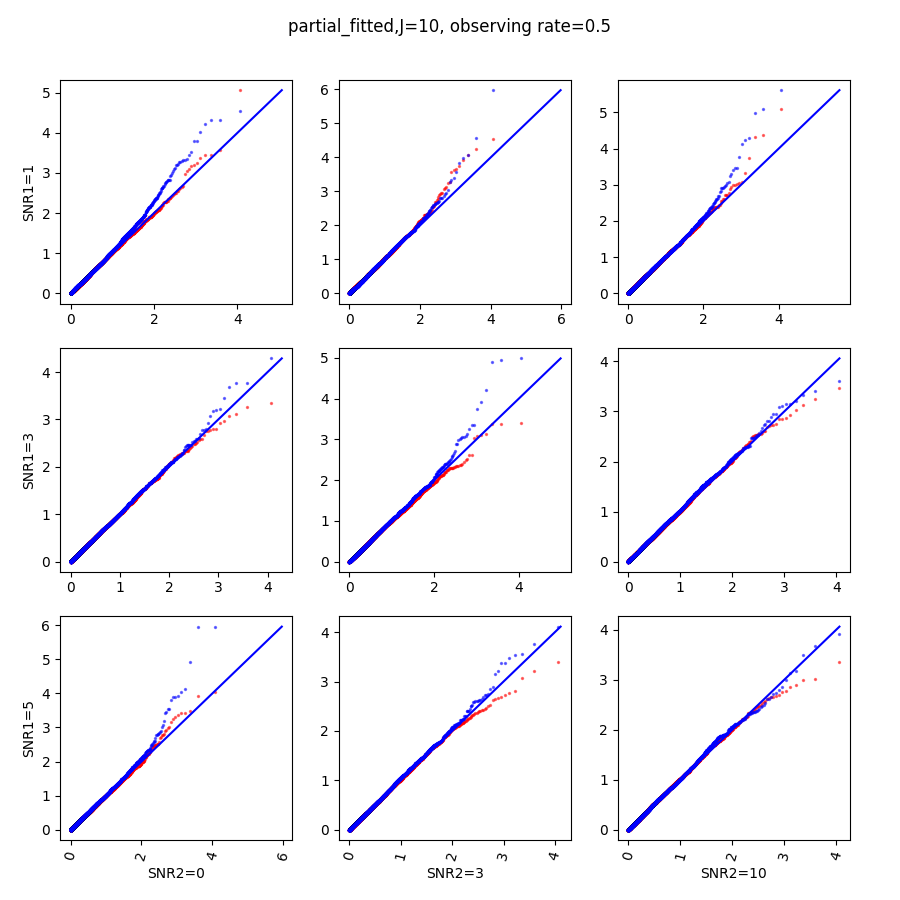}
\includegraphics[width  =0.49\textwidth]{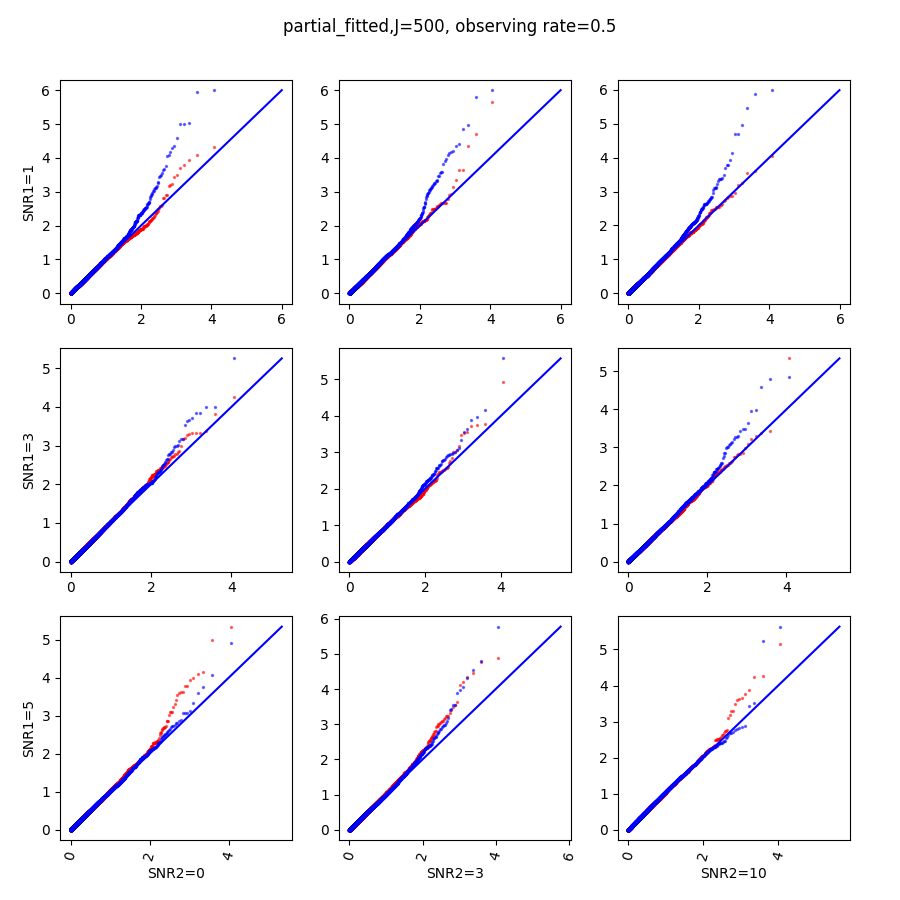}
\end{center}
\end{figure}

\begin{figure}[H]
\caption{qq-plots of constructed p-values at $r = 0.1$. The p values are estimated with nct distribution. The red/blue points represent those from the cross-fit and the naive procedure, with blue diagonal representing the expected theoretical behavior.}
\label{fig:qq_crossfitIII}
\begin{center}
\includegraphics[width  =0.49\textwidth]{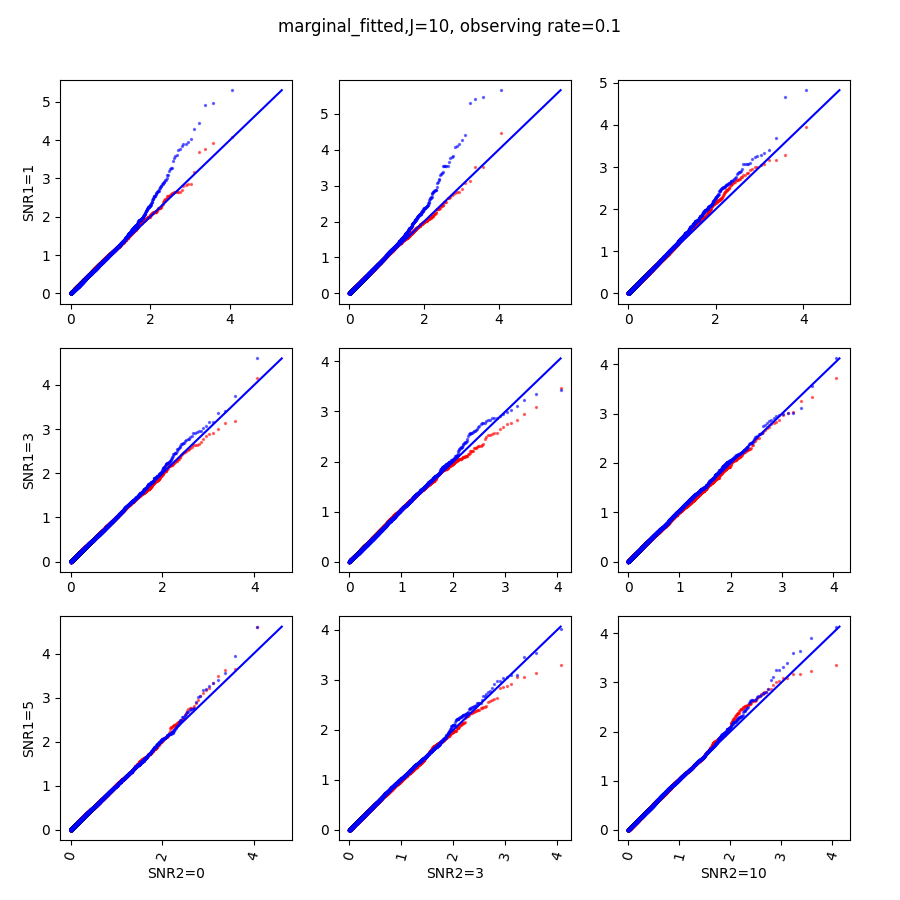}
\includegraphics[width  =0.49\textwidth]{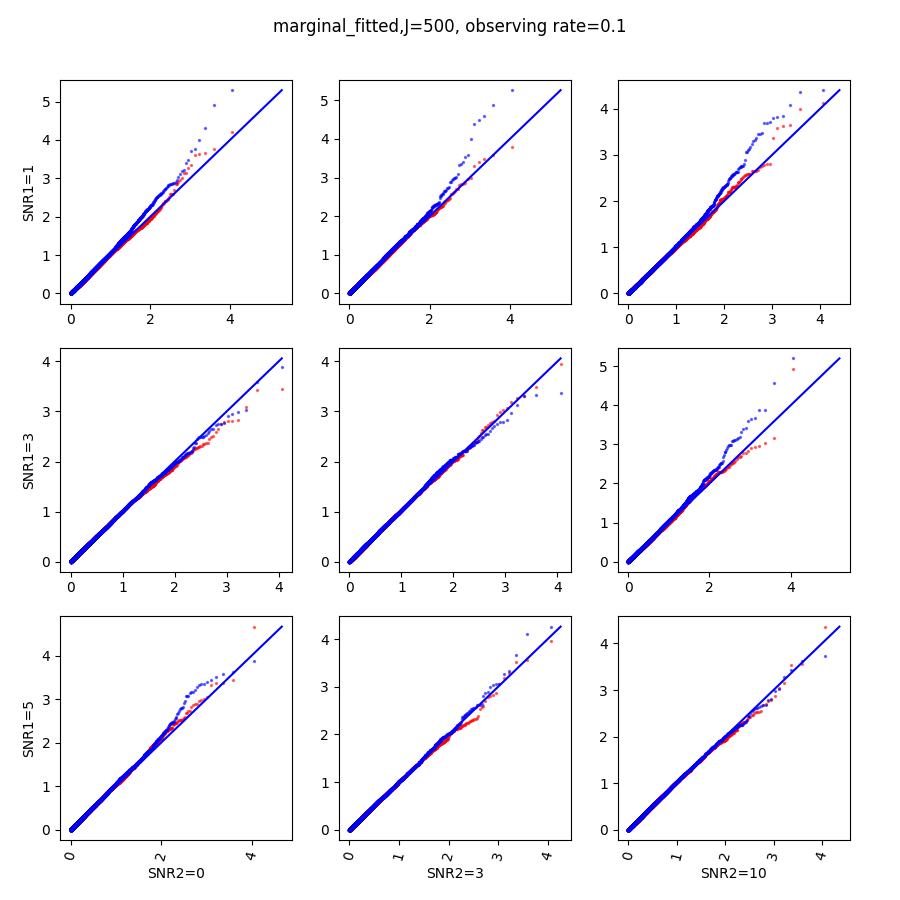}\\
\includegraphics[width  =0.49\textwidth]{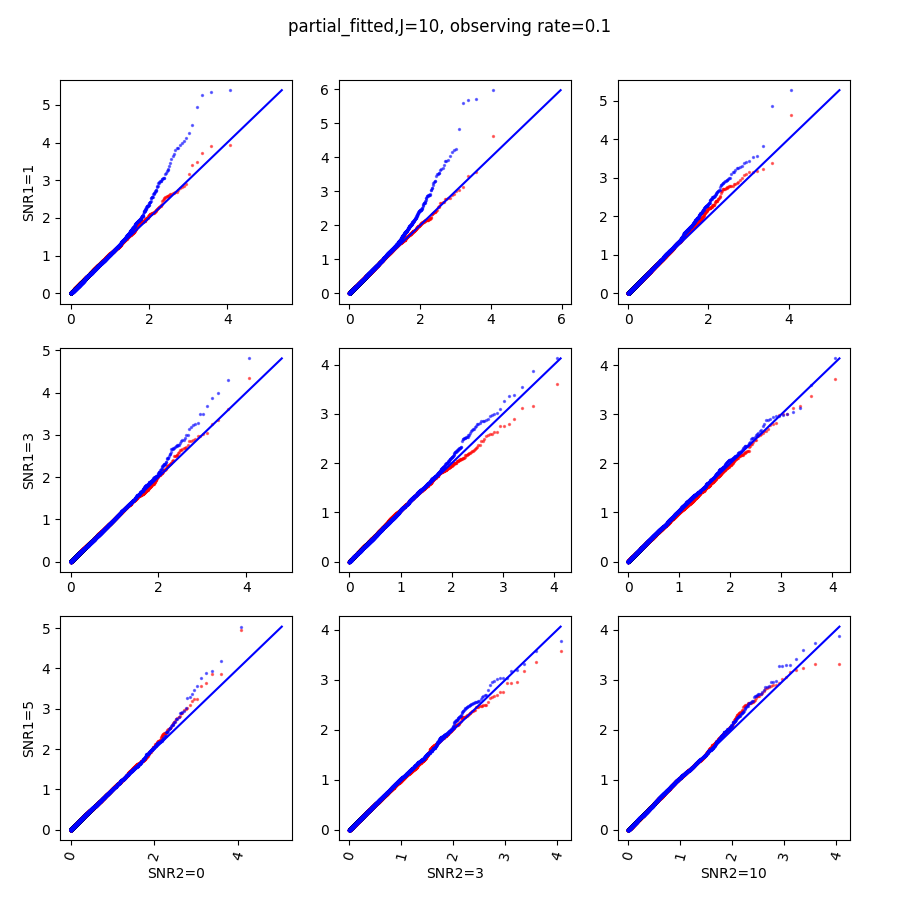}
\includegraphics[width  =0.49\textwidth]{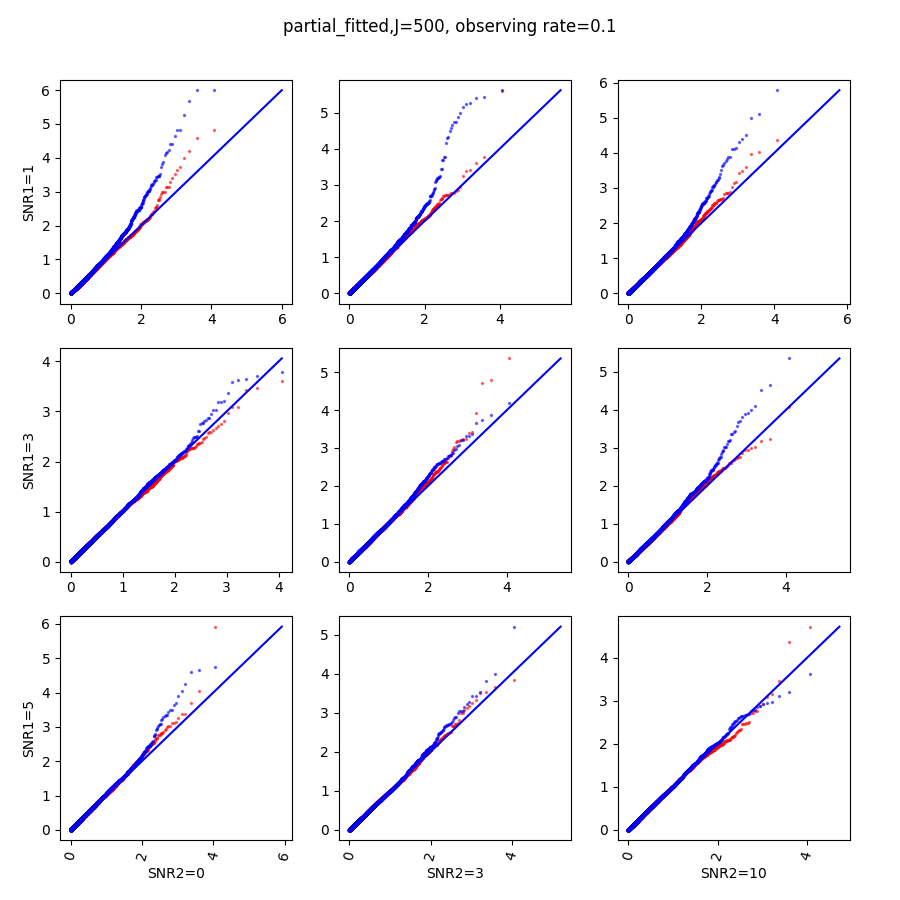}
\end{center}
\end{figure}

In this section, we show qq-plots of the negative $\log_{10}(\rm{p value})$ from the null hypotheses against its theoretical values from the uniform distribution.  In Fig.\ref{fig:qq_crossfitI}-Fig.\ref{fig:qq_crossfitIII}, we show the results from observing rate $r = 1.0, 0.5, 0.1$ respectively, and the red/blue points represent those from the cross fit and the full fit with blue diagonal representing the expected theoretical behavior. The sub-title indicates the dimensionality $J$, observing rate $r\in \{1.0, 0.5, 0.1\}$ and the type of estimate: partial\_fitted means p values from partial independence test with p-value estimated using nct distribution and  marginal\_fitted means p values from marginal independence test with p-value estimated using nct distribution.  The number of randomization used here is $B = 200$. We observe that direct plug-in of model parameters from full fit leads to poor performance when the signal-to-noise ratio is low. When SNR1$=1$, the log pvalue is severely inflated based on the full fit, the cross-fit provides much uniform null p-value distribution even with only a five-step update.

\begin{figure}[H]
\caption{qq-plots of constructed p-values at $r = 1.0$. The p values are estimated directly from the empirical distribution. The red/blue points represent those from the cross-fit and the naive procedure, with blue diagonal representing the expected theoretical behavior.}
\label{fig:qq_nct}
\begin{center}
\includegraphics[width  =0.49\textwidth]{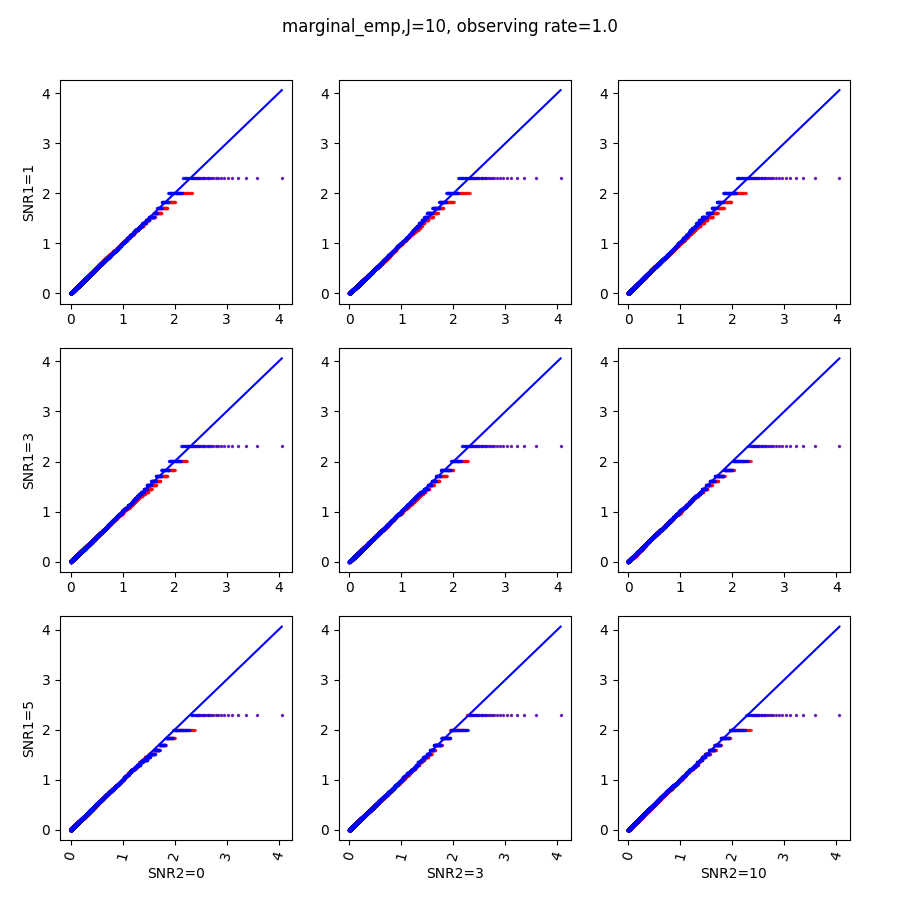}
\includegraphics[width  =0.49\textwidth]{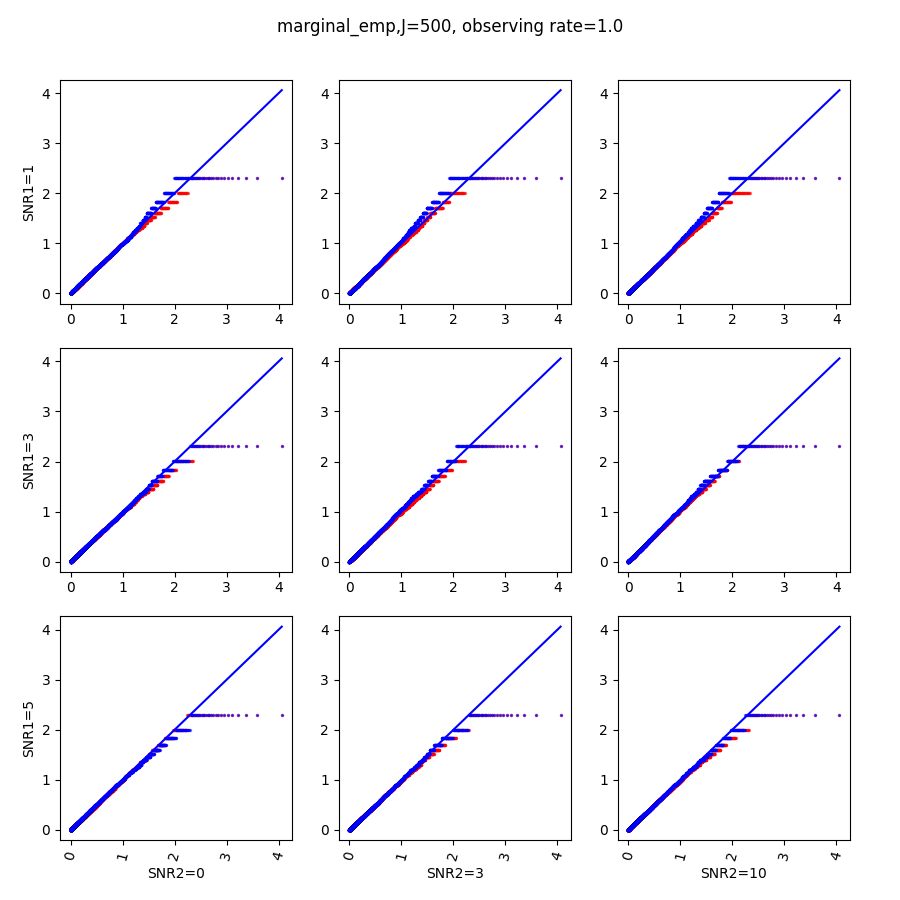}\\
\includegraphics[width  =0.49\textwidth]{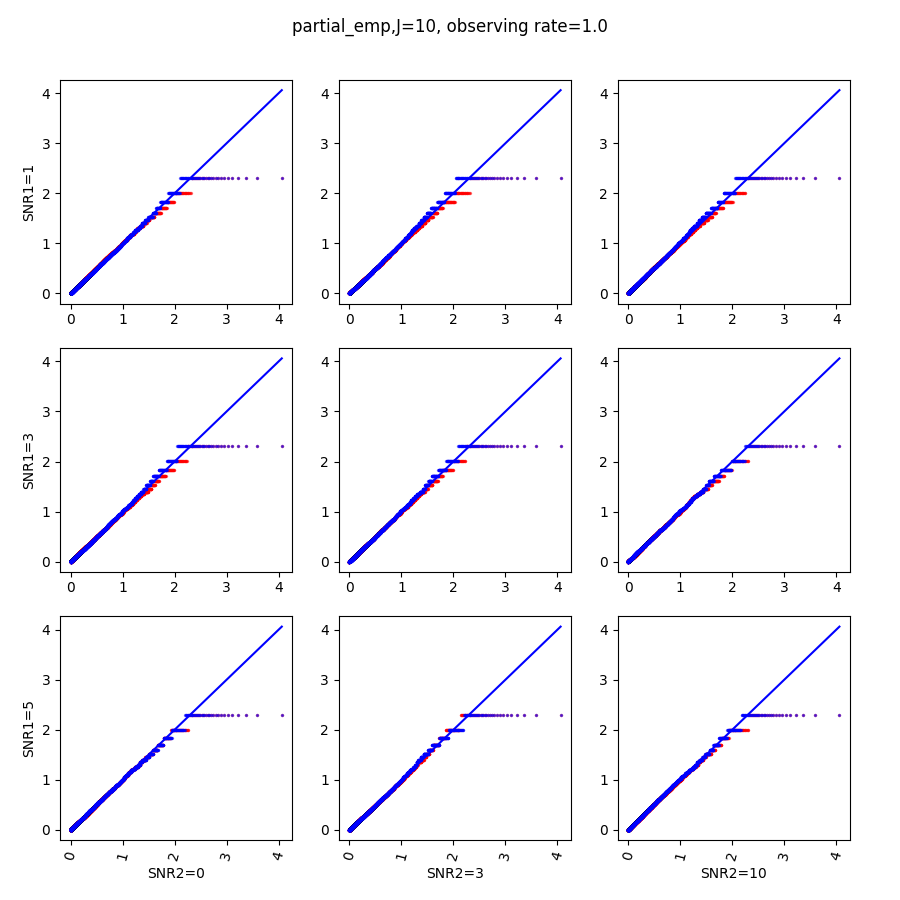}
\includegraphics[width  =0.49\textwidth]{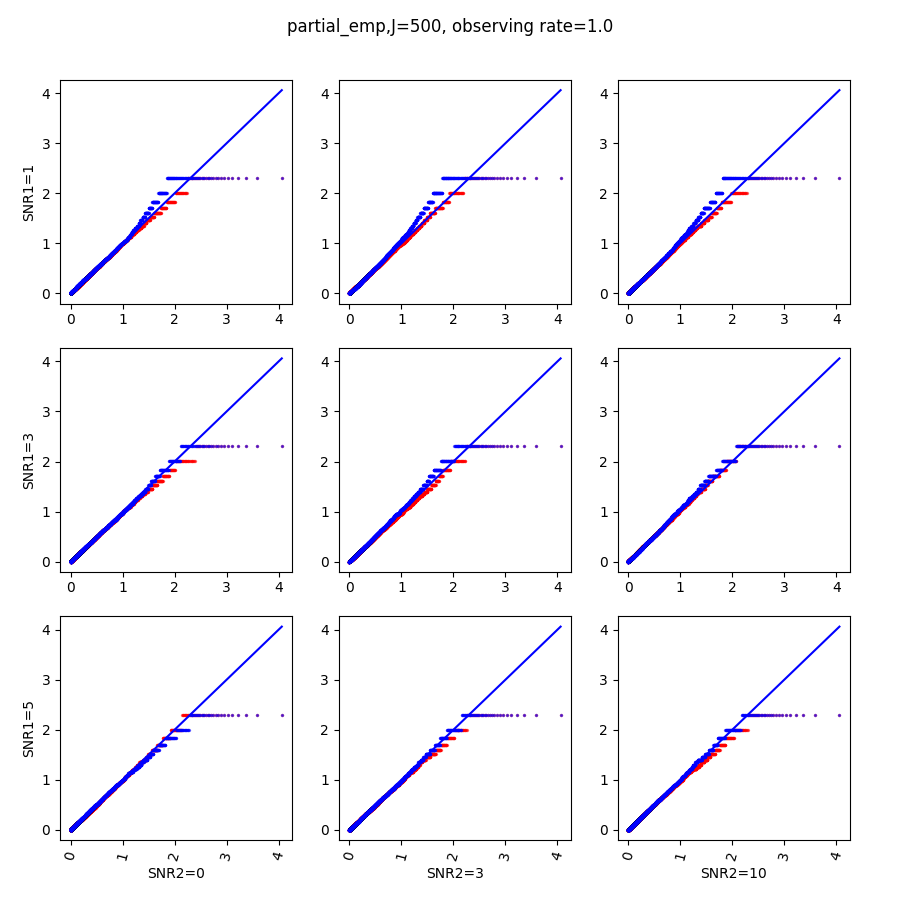}
\end{center}
\end{figure}

When $B = 200$, the non-parametric estimate of the p values has a lower bound of $\frac{1}{B+1}=\frac{1}{201}$, but the nct-estimated p values do not have such a restriction. Fig.\ref{fig:qq_nct} shows the non-parametric p value estimates at $r = 1.0$, we can easily show that the nct-estimated version provides more details for small p-values.
\bibliographystyle{Chicago}
\bibliography{spaco}

\begin{thebibliography}{}

\bibitem[\protect\citeauthoryear{Acar and Yener}{Acar and
  Yener}{2008}]{acar2008unsupervised}
Acar, E. and B.~Yener (2008).
\newblock Unsupervised multiway data analysis: A literature survey.
\newblock {\em IEEE transactions on knowledge and data engineering\/}~{\em
  21\/}(1), 6--20.

\bibitem[\protect\citeauthoryear{Argelaguet, Velten, Arnol, Dietrich, Zenz,
  Marioni, Buettner, Huber, and Stegle}{Argelaguet
  et~al.}{2018}]{argelaguet2018multi}
Argelaguet, R., B.~Velten, D.~Arnol, S.~Dietrich, T.~Zenz, J.~C. Marioni,
  F.~Buettner, W.~Huber, and O.~Stegle (2018).
\newblock Multi-omics factor analysis—a framework for unsupervised
  integration of multi-omics data sets.
\newblock {\em Molecular systems biology\/}~{\em 14\/}(6), e8124.

\bibitem[\protect\citeauthoryear{Bai and Wang}{Bai and
  Wang}{2016}]{bai2016econometric}
Bai, J. and P.~Wang (2016).
\newblock Econometric analysis of large factor models.
\newblock {\em Annual Review of Economics\/}~{\em 8}, 53--80.

\bibitem[\protect\citeauthoryear{Besse and Ramsay}{Besse and
  Ramsay}{1986}]{besse1986principal}
Besse, P. and J.~O. Ramsay (1986).
\newblock Principal components analysis of sampled functions.
\newblock {\em Psychometrika\/}~{\em 51\/}(2), 285--311.

\bibitem[\protect\citeauthoryear{Bro and Andersson}{Bro and
  Andersson}{1998}]{bro1998improving}
Bro, R. and C.~A. Andersson (1998).
\newblock Improving the speed of multiway algorithms: Part ii: Compression.
\newblock {\em Chemometrics and intelligent laboratory systems\/}~{\em
  42\/}(1-2), 105--113.

\bibitem[\protect\citeauthoryear{Cand{\`e}s, Fan, Janson, and Lv}{Cand{\`e}s
  et~al.}{2018}]{candes2018panning}
Cand{\`e}s, E., Y.~Fan, L.~Janson, and J.~Lv (2018).
\newblock Panning for gold:‘model-x’knockoffs for high dimensional
  controlled variable selection series b statistical methodology.

\bibitem[\protect\citeauthoryear{Carroll, Pruzansky, and Kruskal}{Carroll
  et~al.}{1980}]{carroll1980candelinc}
Carroll, J.~D., S.~Pruzansky, and J.~B. Kruskal (1980).
\newblock Candelinc: A general approach to multidimensional analysis of
  many-way arrays with linear constraints on parameters.
\newblock {\em Psychometrika\/}~{\em 45\/}(1), 3--24.

\bibitem[\protect\citeauthoryear{De~Lathauwer, De~Moor, and
  Vandewalle}{De~Lathauwer et~al.}{2000}]{de2000multilinear}
De~Lathauwer, L., B.~De~Moor, and J.~Vandewalle (2000).
\newblock A multilinear singular value decomposition.
\newblock {\em SIAM journal on Matrix Analysis and Applications\/}~{\em
  21\/}(4), 1253--1278.

\bibitem[\protect\citeauthoryear{Dempster, Laird, and Rubin}{Dempster
  et~al.}{1977}]{dempster1977maximum}
Dempster, A.~P., N.~M. Laird, and D.~B. Rubin (1977).
\newblock Maximum likelihood from incomplete data via the em algorithm.
\newblock {\em Journal of the Royal Statistical Society: Series B
  (Methodological)\/}~{\em 39\/}(1), 1--22.

\bibitem[\protect\citeauthoryear{Fan, Fan, and Lv}{Fan
  et~al.}{2008}]{fan2008high}
Fan, J., Y.~Fan, and J.~Lv (2008).
\newblock High dimensional covariance matrix estimation using a factor model.
\newblock {\em Journal of Econometrics\/}~{\em 147\/}(1), 186--197.

\bibitem[\protect\citeauthoryear{Fan, Liao, and Mincheva}{Fan
  et~al.}{2011}]{fan2011high}
Fan, J., Y.~Liao, and M.~Mincheva (2011).
\newblock High dimensional covariance matrix estimation in approximate factor
  models.
\newblock {\em Annals of statistics\/}~{\em 39\/}(6), 3320.

\bibitem[\protect\citeauthoryear{Harshman and Lundy}{Harshman and
  Lundy}{1994}]{harshman1994parafac}
Harshman, R.~A. and M.~E. Lundy (1994).
\newblock Parafac: Parallel factor analysis.
\newblock {\em Computational Statistics \& Data Analysis\/}~{\em 18\/}(1),
  39--72.

\bibitem[\protect\citeauthoryear{Hinrich and M{\o}rup}{Hinrich and
  M{\o}rup}{2019}]{hinrich2019probabilistic}
Hinrich, J.~L. and M.~M{\o}rup (2019).
\newblock Probabilistic tensor train decomposition.
\newblock In {\em 2019 27th European Signal Processing Conference (EUSIPCO)},
  pp.\  1--5. IEEE.

\bibitem[\protect\citeauthoryear{Huang, Shen, Buja, et~al.}{Huang
  et~al.}{2008}]{huang2008functional}
Huang, J.~Z., H.~Shen, A.~Buja, et~al. (2008).
\newblock Functional principal components analysis via penalized rank one
  approximation.
\newblock {\em Electronic Journal of Statistics\/}~{\em 2}, 678--695.

\bibitem[\protect\citeauthoryear{Imaizumi and Hayashi}{Imaizumi and
  Hayashi}{2017}]{imaizumi2017tensor}
Imaizumi, M. and K.~Hayashi (2017).
\newblock Tensor decomposition with smoothness.
\newblock In {\em International Conference on Machine Learning}, pp.\
  1597--1606. PMLR.

\bibitem[\protect\citeauthoryear{Katsevich and Roeder}{Katsevich and
  Roeder}{2020}]{katsevich2020conditional}
Katsevich, E. and K.~Roeder (2020).
\newblock Conditional resampling improves sensitivity and specificity of single
  cell crispr regulatory screens.
\newblock {\em bioRxiv\/}.

\bibitem[\protect\citeauthoryear{Lam, Yao, and Bathia}{Lam
  et~al.}{2011}]{lam2011estimation}
Lam, C., Q.~Yao, and N.~Bathia (2011).
\newblock Estimation of latent factors for high-dimensional time series.
\newblock {\em Biometrika\/}~{\em 98\/}(4), 901--918.

\bibitem[\protect\citeauthoryear{Li, Shen, and Huang}{Li
  et~al.}{2016}]{li2016supervised}
Li, G., H.~Shen, and J.~Z. Huang (2016).
\newblock Supervised sparse and functional principal component analysis.
\newblock {\em Journal of Computational and Graphical Statistics\/}~{\em
  25\/}(3), 859--878.

\bibitem[\protect\citeauthoryear{Lock and Li}{Lock and
  Li}{2018}]{lock2018supervised}
Lock, E.~F. and G.~Li (2018).
\newblock Supervised multiway factorization.
\newblock {\em Electronic journal of statistics\/}~{\em 12\/}(1), 1150.

\bibitem[\protect\citeauthoryear{Lucas, Wong, Klein, Castro, Silva, Sundaram,
  Ellingson, Mao, Oh, Israelow, et~al.}{Lucas
  et~al.}{2020}]{lucas2020longitudinal}
Lucas, C., P.~Wong, J.~Klein, T.~B. Castro, J.~Silva, M.~Sundaram, M.~K.
  Ellingson, T.~Mao, J.~E. Oh, B.~Israelow, et~al. (2020).
\newblock Longitudinal analyses reveal immunological misfiring in severe
  covid-19.
\newblock {\em Nature\/}~{\em 584\/}(7821), 463--469.

\bibitem[\protect\citeauthoryear{Mnih and Salakhutdinov}{Mnih and
  Salakhutdinov}{2007}]{mnih2007probabilistic}
Mnih, A. and R.~R. Salakhutdinov (2007).
\newblock Probabilistic matrix factorization.
\newblock {\em Advances in neural information processing systems\/}~{\em 20},
  1257--1264.

\bibitem[\protect\citeauthoryear{Phan, Tichavsk{\`y}, and Cichocki}{Phan
  et~al.}{2013}]{phan2013candecomp}
Phan, A.-H., P.~Tichavsk{\`y}, and A.~Cichocki (2013).
\newblock Candecomp/parafac decomposition of high-order tensors through tensor
  reshaping.
\newblock {\em IEEE transactions on signal processing\/}~{\em 61\/}(19),
  4847--4860.

\bibitem[\protect\citeauthoryear{Rendeiro, Casano, Vorkas, Singh, Morales,
  DeSimone, Ellsworth, Soave, Kapadia, Saito, et~al.}{Rendeiro
  et~al.}{2020}]{rendeiro2020longitudinal}
Rendeiro, A.~F., J.~Casano, C.~K. Vorkas, H.~Singh, A.~Morales, R.~A. DeSimone,
  G.~B. Ellsworth, R.~Soave, S.~N. Kapadia, K.~Saito, et~al. (2020).
\newblock Longitudinal immune profiling of mild and severe covid-19 reveals
  innate and adaptive immune dysfunction and provides an early prediction tool
  for clinical progression.
\newblock {\em medRxiv\/}.

\bibitem[\protect\citeauthoryear{Sidiropoulos, De~Lathauwer, Fu, Huang,
  Papalexakis, and Faloutsos}{Sidiropoulos
  et~al.}{2017}]{sidiropoulos2017tensor}
Sidiropoulos, N.~D., L.~De~Lathauwer, X.~Fu, K.~Huang, E.~E. Papalexakis, and
  C.~Faloutsos (2017).
\newblock Tensor decomposition for signal processing and machine learning.
\newblock {\em IEEE Transactions on Signal Processing\/}~{\em 65\/}(13),
  3551--3582.

\bibitem[\protect\citeauthoryear{Sorkine, Cohen-Or, Lipman, Alexa, R{\"o}ssl,
  and Seidel}{Sorkine et~al.}{2004}]{sorkine2004laplacian}
Sorkine, O., D.~Cohen-Or, Y.~Lipman, M.~Alexa, C.~R{\"o}ssl, and H.-P. Seidel
  (2004).
\newblock Laplacian surface editing.
\newblock In {\em Proceedings of the 2004 Eurographics/ACM SIGGRAPH symposium
  on Geometry processing}, pp.\  175--184.

\bibitem[\protect\citeauthoryear{Tibshirani}{Tibshirani}{2011}]{tibshirani2011regression}
Tibshirani, R. (2011).
\newblock Regression shrinkage and selection via the lasso: a retrospective.
\newblock {\em Journal of the Royal Statistical Society: Series B (Statistical
  Methodology)\/}~{\em 73\/}(3), 273--282.

\bibitem[\protect\citeauthoryear{Tipping and Bishop}{Tipping and
  Bishop}{1999}]{tipping1999probabilistic}
Tipping, M.~E. and C.~M. Bishop (1999).
\newblock Probabilistic principal component analysis.
\newblock {\em Journal of the Royal Statistical Society: Series B (Statistical
  Methodology)\/}~{\em 61\/}(3), 611--622.

\bibitem[\protect\citeauthoryear{Wang, Liu, and Chen}{Wang
  et~al.}{2019}]{wang2019factor}
Wang, D., X.~Liu, and R.~Chen (2019).
\newblock Factor models for matrix-valued high-dimensional time series.
\newblock {\em Journal of econometrics\/}~{\em 208\/}(1), 231--248.

\bibitem[\protect\citeauthoryear{Wang, Zheng, Lian, and Li}{Wang
  et~al.}{2021}]{wang2021high}
Wang, D., Y.~Zheng, H.~Lian, and G.~Li (2021).
\newblock High-dimensional vector autoregressive time series modeling via
  tensor decomposition.
\newblock {\em Journal of the American Statistical Association\/}, 1--19.

\bibitem[\protect\citeauthoryear{Yao, M{\"u}ller, and Wang}{Yao
  et~al.}{2005}]{yao2005functional}
Yao, F., H.-G. M{\"u}ller, and J.-L. Wang (2005).
\newblock Functional data analysis for sparse longitudinal data.
\newblock {\em Journal of the American statistical association\/}~{\em
  100\/}(470), 577--590.

\bibitem[\protect\citeauthoryear{Yokota, Zhao, and Cichocki}{Yokota
  et~al.}{2016}]{yokota2016smooth}
Yokota, T., Q.~Zhao, and A.~Cichocki (2016).
\newblock Smooth parafac decomposition for tensor completion.
\newblock {\em IEEE Transactions on Signal Processing\/}~{\em 64\/}(20),
  5423--5436.

\end{thebibliography}
\end{document}